\documentclass[letter,longauth]{aa} % for the letters 
\usepackage{natbib}
\bibpunct{(}{)}{;}{a}{}{,} % to follow the A&A style
\usepackage{booktabs}
\usepackage{verbatim,multirow,multicol}
\usepackage{subfig}
\usepackage{hhline}
\usepackage{soul}
\setstcolor{red}
\usepackage{color}
\usepackage{tablefootnote}

\definecolor{Red}{rgb}{0.9,0,0}
\definecolor{Blue}{rgb}{0,0,0.9}
\definecolor{Green}{rgb}{0,0.5,0}
\definecolor{Black}{rgb}{0,0,0}

\newcommand{\Gaia}{\textit{Gaia}~}

\newcommand\bla{\color{Black}}

\usepackage{txfonts}
\usepackage{graphicx}
\usepackage{multirow}
\usepackage{booktabs}
\usepackage{txfonts}
\usepackage{hyperref}

\begin{document} 

\title{The two rings of (50000) Quaoar}
\author{
C.~L.~Pereira\inst{1,2}\thanks{\email{chrystianpereira@on.br}}%,orcid: 0000-0003-1000-8113
\and
B.~Sicardy\inst{3} % email: bruno.sicardy@obspm.fr,orcid: 0000-0003-1995-0842
\and
B.~E.~Morgado\inst{4,1,2} % email: morgado.fis@gmail.com,orcid: 0000-0003-0088-1808
\and
F.~Braga-Ribas\inst{5,1,2} % email: felipebribas@gmail.com,orcid: 0000-0003-2311-2438
\and
E.~Fern\'{a}ndez-Valenzuela\inst{6,7} % email: estela@ucf.edu,orcid: 0000-0003-2132-7769
\and
D.~Souami\inst{3,8,9,10}\thanks{Fulbright Visiting Scholar (2022 - 2023) at University of California, Berkeley} % email: damya.souami@obspm.fr,orcid: 0000-0003-4058-0815
\and
B.~J.~Holler\inst{11} % email: bholler@stsci.edu,orcid: 0000-0002-6117-0164
\and
R.~C.~Boufleur\inst{1,2} % email: rcboufleur@gmail.com ,orcid: 0000-0003-3452-1114
\and
G.~Margoti\inst{5} % email: giulianomargoti@alunos.utfpr.edu.br,
\and
M.~Assafin\inst{4,2} % email: massaf@ov.ufrj.br,orcid: 0000-0002-8211-0777
\and
J.~L.~Ortiz\inst{7} % email: ortiz@iaa.es,orcid: 0000-0002-8690-2413
\and
P.~Santos-Sanz\inst{7} % email: psantos@iaa.es,orcid: 0000-0002-1123-9830
\and
B.~Epinat\inst{12,13} % email: benoit.epinat@lam.fr,
\and
P.~Kervella\inst{3} % email: pierre.kervella@obspm.fr,orcid: 0000-0003-0626-1749
\and
J.~Desmars\inst{14,15} % email: josselin.desmars@obspm.fr,orcid: 0000-0002-2193-8204
\and
R.~Vieira-Martins\inst{1,2} % email: rvm@on.br,orcid: 0000-0003-1690-5704
\and
Y.~Kilic\inst{16,17} % email: yucelkilic1@gmail.com,orcid: 0000-0001-8641-0796
\and
A.~R.~Gomes-J\'{u}nior\inst{18,19,2} % email: altairgomesjr@gmail.com,orcid: 0000-0002-3362-2127
\and
J.~I.~B.~Camargo\inst{1,2} % email: camargo@on.br,
\and
M.~Emilio\inst{20,1,5} % email: marcelo_emilio@yahoo.com,orcid: 0000-0001-5589-9015
\and
M.~Vara-Lubiano\inst{7} % email: mvara@iaa.es,orcid: 0000-0002-8112-0770
\and
M.~Kretlow\inst{7,21,22} % email: mike@kretlow.de,orcid: 0000-0001-8858-3420
\and
L.~Albert\inst{23,24} % email: loic.albert@umontreal.ca,orcid: 0000-0003-0475-9375
\and
C.~Alcock\inst{25} % 
\and
J.~G.~Ball\inst{26} % email: jesse.ball@noirlab.edu,
\and
K.~Bender\inst{27} % email: umetempura@hotmail.com,
\and
M.~W.~Buie\inst{28} % email: buie@boulder.swri.edu,
\and
K.~Butterfield\inst{29} % 
\and
M.~Camarca\inst{30} % email: mcamarca@caltech.edu,
\and
J.~H.~Castro-Chac\'{o}n\inst{31} % 
\and
R.~Dunford\inst{29} % email: bargonne@gmail.com,
\and
R.~S.~Fisher\inst{32} % email: rsf@uoregon.edu,orcid: 0000-0002-9643-7543
\and
D.~Gamble\inst{29,33} % 
\and
J.~C.~Geary\inst{25} % 
\and
C.~L.~Gnilka\inst{34} % email: clgnilka@gmail.com,
\and
K.~D.~Green\inst{35} % email: kgreen@was-ct.org,
\and
Z.~D.~Hartman\inst{26} % email: zachary.hartman@noirlab.edu,orcid: 0000-0003-4236-6927
\and
C-K.~Huang\inst{36} % 
\and
H.~Januszewski\inst{13} % email: helenj@cfht.hawaii.edu,
\and
J.~Johnston\inst{37} % email: Julia.Johnston@colorado.edu,
\and
M.~Kagitani\inst{38} % email: masato.kagitani.c7@tohoku.ac.jp,orcid: 0000-0002-4115-7122
\and
R.~Kamin\inst{29} % email: rlkamin@prodigy.net,
\and
J.~J.~Kavelaars\inst{39} % 
\and
J.~M.~ Keller\inst{37} % email: John.M.Keller@colorado.edu,
\and
K.~R.~de~Kleer\inst{30} % email: dekleer@caltech.edu,
\and
M.~J.~Lehner\inst{36} % email: mlehner@asiaa.sinica.edu.tw,
\and
A.~Luken\inst{32} % email: aluken@uoregon.edu,
\and
F.~Marchis\inst{40,41} % email: fmarchis@seti.org,orcid: 0000-0001-7016-7277
\and
T.~Marlin\inst{30} % email: kmcgregor@wesleyan.edu,
\and
K.~McGregor\inst{42} % email: kmcgregor@wesleyan.edu,orcid: 0000-0003-2111-3437
\and
V.~Nikitin\inst{29,43} % email: nikitin@rocketmail.com,
\and
R.~Nolthenius\inst{27} % email: rickn27@hotmail.com,
\and
C.~Patrick\inst{29} % 
\and
S.~Redfield\inst{42} % email: sredfield@wesleyan.edu,orcid: 0000-0003-3786-3486
\and
A.~W.~Rengstorf\inst{44} % email: adamwr@pnw.edu,
\and
M.~Reyes-Ruiz\inst{31} % 
\and
T.~Seccull\inst{26} % email: tom.seccull@noirlab.edu,orcid: 0000-0001-5605-1702
\and
M.~F.~Skrutskie\inst{45} % email: mfs4n@virginia.edu,
\and
A.~B.~Smith\inst{26} % email: adam.smith@noirlab.edu,orcid: 0000-0002-5250-0045
\and
M.~Sproul\inst{46} % email: msproul@skychariot.com,
\and
A.~W.~Stephens\inst{26} % email: andrew.stephens@noirlab.edu,orcid: 0000-0002-4434-2307
\and
A.~Szentgyorgyi\inst{25} % 
\and
S.~S\'{a}nchez-Sanjuán\inst{31} % 
\and
E.~Tatsumi\inst{47,48} % email: etatsumi@iac.es,
\and
A.~Verbiscer\inst{45} % email: verbiscer@boulder.swri.edu,
\and
S-Y.~Wang\inst{36} % 
\and
F.~Yoshida\inst{49,50} % email: fumi.yoshida.ermei@gmail.com,
\and
R.~Young\inst{51} % email: youngbob2@aim.com,
\and
Z-W.~Zhang\inst{36} % 
}

\institute{
Observat\'{o}rio Nacional/MCTI, R. General Jos\'{e} Cristino 77, CEP 20921-400 Rio de Janeiro - RJ, Brazil \\
\email{chrystianpereira@on.br}
\and
Laborat\'{o}rio Interinstitucional de e-Astronomia - LIneA, Rio de Janeiro, RJ, Brazil
\and
LESIA, Observatoire de Paris, Universit\'{e} PSL, Sorbonne Universit\'{e}, Universit\'{e} de Paris, CNRS, 92190 Meudon, France
\and
Universidade Federal do Rio de Janeiro - Observat\'{o}rio do Valongo, Ladeira Pedro Antônio 43, CEP 20.080-090, Rio de Janeiro - RJ, Brazil
\and
Federal University of Technology - Paran\'{a} (UTFPR-Curitiba), Rua Sete de Setembro, 3165, CEP 80230-901, Curitiba, PR, Brazil
\and
Florida Space Institute, UCF, 12354 Research Parkway, Partnership 1 building, Room 211, Orlado, USA
\and
Instituto de Astrofísica de Andaluc\'{i}a – Consejo Superior de Investigaciones Cient\'{i}ficas, Glorieta de la Astronom\'{i}a S/N, E-18008, Granada, Spain
\and
Departments of Astronomy, and of Earth and Planetary Science, 501, Campbell Hall, University of California, Berkeley, CA 94720, USA
\and
naXys, Department of Mathematics, University of Namur, Rue de Bruxelles 61, 5000 Namur, Belgium
\and
Universit\'{e} C\^{o}te d’Azur, Observatoire de la C\^{o}te d’Azur, CNRS, Laboratoire Lagrange, Bd de l’Observatoire, CS 34229, 06304 Nice Cedex 4, France
\and
Space Telescope Science Institute, Baltimore, Maryland, USA
\and
Aix Marseille Universit\'{e}, CNRS, CNES, LAM (Laboratoire d’Astrophysique de Marseille) UMR 7326, 13388, Marseille, France
\and
Canada-France-Hawaii Telescope, 65-1238 Mamalahoa Highway, Kamuela, HI 96743, USA
\and
Institut Polytechnique des Sciences Avanc\'{e}es IPSA, 63 boulevard de Brandebourg, 94200 Ivry-sur-Seine, France
\and
Institut de M\'{e}canique C\'{e}leste et de Calcul des \'{E}ph\'{e}m\'{e}rides, IMCCE, Observatoire de Paris, PSL Research University, CNRS, Sorbonne Universit\'{e}s, UPMC Univ Paris 06, Univ. Lille, France
\and
Akdeniz University, Faculty of Sciences, Department of Space Sciences and Technologies, 07058 Antalya, Turkey
\and
TÜB\.{I}TAK National Observatory, Akdeniz University Campus, 07058 Antalya, Turkey
\and
Institute of Physics, Federal University of Uberl\^{a}ndia, Uberl\^{a}ndia-MG, Brazil
\and
UNESP - S\~{a}o Paulo State University, Grupo de Din\^{a}mica Orbital e Planetologia, CEP 12516-410, Guaratinguet\'{a}, SP, Brazil
\and
Universidade Estadual de Ponta Grossa, O.A. - DEGEO, Ponta Grossa (PR), Brazil
\and
Internationale Amateursternwarte (IAS) e. V., Mittelstr. 6, 15749 Mittenwalde, Germany
\and
International Occultation Timing Association - European Section (IOTA/ES), Am Brombeerhag 13, 30459 Hannover, Germany
\and
D\'{e}partement de Physique and Observatoire du Mont-M\'{e}gantic, Universit\'e de Montr\'{e}al, C.P. 6128, Succ. Centre-ville, Montr\'{e}al, H3C 3J7, Qu\'{e}bec, Canada
\and
Institut Trottier de Recherche sur les exoplan\`{e}tes, Universit\'{e} de Montr\'{e}al
\and
Harvard-Smithsonian Center for Astrophysics, 60 Garden Street, Cambridge, MA 02138, USA
\and
Gemini Observatory/NSF’s NOIRLab, Hilo, Hawaii, USA
\and
Cabrillo College Astronomy, USA
\and
Southwest Research Institute, 1050 Walnut Street, Suite 300, Boulder, CO 80302, USA
\and
International Occultation Timing Association (IOTA), USA
\and
California Institute of Technology, 1200 E California Blvd, Pasadena, CA 91125, USA
\and
Universidad Nacional Aut\'onoma de M\'exico, Instituto de Astronom\'ia, AP 106, Ensenada, 22800, BC, Mexico
\and
University of Oregon, Eugene, OR, USA
\and
CanCON - Canadian Collaborative Occultation Network, Canada
\newpage %Add here by BEM
\and
NASA Ames Research Center, Moffett Field, California, USA \& NASA Exoplanet Science Institute, Caltech/IPAC, Mail Code 100-22, Pasadena, CA, USA
\and
University of New Haven, Dept. of Mathematics and Physics, 300 Boston Post Road, West Haven, CT 06477
\and
Academia Sinica Institute of Astronomy and Astrophysics, 11F of AS/NTU Astronomy-Mathematics Building, No.1, Sec. 4, Roosevelt Road, Taipei 10617, Taiwan, R.O.C.
\and
University of Colorado, 2000 Colorado Avenue, Boulder, CO 80309, USA
\and
Planetary Plasma and Atmospheric Research Center, Graduate School of Science, Tohoku University, 6-3 Aramaki -aza-aoba, Aoba-ku, Sendai 980-8578, Japan
\and
Department of Physics and Astronomy, University of Victoria, Building, 3800 Finnerty Road, Victoria, BC V8P 5C2, Canada
\and
Unistellar, 5 all\'ee Marcel Leclerc, bâtiment B, 13008 Marseille, France
\and
SETI Institute, Carl Sagan Center, Suite 200, 339 Bernardo Avenue, Mountain View, CA 94043, USA
\and
Astronomy Department and Van Vleck Observatory, Wesleyan University, Middletown, CT 06459, USA
\and
Research and Education Collaborative Occultation Network, USA
\and
Purdue University Northwest, Department of Chemistry and Physics, Hammond, IN, USA
\and
University of Virginia, Department of Astronomy, P.O. Box 400325, Charlottesville, VA 22904, USA
\and
Private Observatory, Pennsylvania, USA
\and
Instituto de Astrofísica de Canarias, University of La Laguna, Tenerife, Spain
\and
Dept. Earth and Planetary Science, The University of Tokyo, Tokyo, Japan
\and
School of Medicine Department of Basic Sciences University of Occupational and Environmental Health, Japan 1-1 Iseigaoka, Yahata, Kitakyusyu 807-8555, JAPAN
\and
Planetary Exploration Research Center, Chiba Institute of Technology 2-17-1 Tsudanuma,Narashino, Chiba 275-0016, Japan
\and
Naylor Observatory
}

\date{
Received 10 March 2023 / Accepted 17 April 2023
}

\abstract
{
Quaoar is a classical trans-Neptunian object (TNO) with an area-equivalent diameter of 1100 km and an orbital semi-major axis of 43.3 astronomical units. Based on stellar occultations observed between 2018 and 2021, an inhomogeneous ring (Q1R, i.e., Quaoar’s first ring) has been detected around this body.
}
{
A new stellar occultation by Quaoar was observed on August 9$^{\rm{}}$, 2022, with the aim of improving Quaoar’s shape models and the physical parameters of Q1R, while searching for additional material around the body.}
{
The occultation provided nine effective chords across Quaoar, pinning down its size, shape, and astrometric position. Large facilities, such as Gemini North and the Canada-France-Hawaii Telescope (CFHT), were used to obtain high acquisition rates and signal-to-noise ratios. The light curves were also used to characterize the Q1R ring (radial profiles and orbital elements).
}
{
Quaoar's elliptical fit to the occultation chords yields the limb with an apparent semi-major axis of $579.5 \pm 4.0$ km, apparent oblateness of $0.12 \pm 0.01$, and area-equivalent radius of $543 \pm 2$ km. Quaoar’s limb orientation is consistent with Q1R and Weywot orbiting in Quaoar’s equatorial plane. The orbital radius of Q1R is refined to a value of $4,057 \pm 6$ km. The radial opacity profile of the more opaque ring profile follows a Lorentzian shape that extends over 60 km, with a full width at half maximum (FWHM) of $\sim$5 km and a peak normal optical depth of 0.4. Besides the secondary events related to the already reported rings, new secondary events detected during the August 2022 occultation in three different data sets are consistent with another ring around Quaoar with a radius of $2,520 \pm 20$ km, assuming the ring is circular and co-planar with Q1R. This new ring has a typical width of 10 km and a normal optical depth of $\sim$0.004. Just as Q1R, it also lies outside Quaoar’s classical Roche limit.
}
{}

\maketitle
%
%-------------------------------------------------------------------
\section{Introduction}
In the last decade, three ring systems have been discovered around minor bodies in the outer Solar System: the Centaur Chariklo \citep{BragaRibas2014}, the dwarf planet Haumea \citep{Ortiz2017}, and the Trans-Neptunian Object (TNO) (50000) Quaoar \citep{Morgado2023}. Dense material has also been detected around the Centaur Chiron \citep{Ruprecht2015, Ortiz2015, Sickafoose2020}. However, the nature of this material, namely, whether it is a permanent or transient ring or a dust shell, is still a matter of debate.

Quaoar's ring, referred to as Q1R hereafter, was detected during several stellar occultations observed between 2018 and 2021 \citep{Morgado2023}.
Q1R has a radius of about 4,100 km with significant azimuthal variations in the optical depth, ranging between 0.004 and 0.1-0.7, and in width, ranging from 5 km to 300 km.
Like Chariklo's and Haumea's rings, Quaoar's Q1R ring orbits close to the 1/3 spin-orbit resonance (SOR) with the central body, suggesting a link between this resonance and the ring \citep{Salo2021EPSC, Sicardy2021EPSC, Morgado2023}. Meanwhile, a unique property of Q1R is its location, which is far outside Quaoar's classical Roche limit. This limit is estimated to be at 1,780 km from the body center, assuming particles with a bulk density of $\rho = 0.4\,\rm{g}\,\rm{cm}^{-3}$. Outside the Roche limit, rings should accrete into satellites over timescales of less than 100 years \citep{Kokubo2000, Takeda2001}. However, collisions more elastic than previously considered for Saturn's ring may maintain a ring unaccreted at distances greater than the Roche limit \citep{Morgado2023}. The 6:1 mean-motion resonance (MMR) of Quaoar and its satellite, Weywot, may contribute to the confinement of the ring since the satellite's eccentricity creates an equilibrium region capable of concentrating ring material in an arc over a longitude interval \citep{Morgado2023}. The confinement of the rings may also be due to the presence of putative "shepherd" satellites.

This letter presents observations of a stellar occultation by Quaoar that occurred on August 9, 2022. This event was observed in Hawaii, the continental US, and Mexico. The high image acquisition rate and high signal-to-noise-ratios (S/Ns) obtained at Mauna Kea with the `Alopeke camera at the Gemini North 8.1-m telescope and Wide-field InfraRed Camera (WIRCam) at the Canada-France-Hawaii 3.6-m Telescope (CFHT) allowed the dense part of Q1R to be radially resolved and its optical depth to be probed in the r', z' and Ks bands. Furthermore, this event revealed a new ring around Quaoar, hereafter called the Q2R (Quaoar's second ring).

%--------------------------------------------------------------------
\section{Prediction and observation}
\label{sec:prediction_observation}
The occultation of August 9, 2022 was predicted within the framework of the European Research Council (ERC) \textit{Lucky Star} project\footnote{\url{https://lesia.obspm.fr/lucky-star}}. The campaign was managed by the \textit{Occultation Portal} website\footnote{\url{https://occultation.tug.tubitak.gov.tr}} described in \cite{Kilic2022}. The occultation shadow crossed the continental United States, Mexico, and the Hawaiian archipelago with a %\st{geocentric closest approach at 06:34:04 UTC and}
geocentric shadow velocity of 17.6~km~s$^{-1}$. The shadow path and the sites involved in the observation campaign are shown in Appendix \ref{fig:occ_map}. More details on the occulted star and Quaoar are provided in Table \ref{tab:star_obj}.

\begin{table}[!ht]
    \setlength{\tabcolsep}{2mm}
    \centering
    \caption{
    More details about the occulted star and Quaoar.
    }
    \begin{tabular}{l l}
    \toprule \toprule
    \multicolumn{2}{c}{Occulted star} \\
    \midrule
    Epoch                              & 2022-08-09 06:34:04 UTC  \\
    Source ID                          & \Gaia DR3 4098214367441486592 \\
    Star position                 & $\alpha_\star$= 18$^h$21$^m$42$^s$.86965 $\pm$ 0.2404 mas   \\
    \hspace{5mm}at epoch\tablefootmark{1}     & $\delta_\star$= -15$^\circ$12$'$45$''$.9639 $\pm$ 0.2191 mas \\
    Magnitudes\tablefootmark{2}        &  G = 15.3; RP = 14.1; BP = 16.8;\\
    &   J = 12.0; H = 11.0; K = 10.7 \\
    Apparent diameter\tablefootmark{3} &  0.44 mas / 1.33 km \\
    RUWE\tablefootmark{4}                          & 0.959      \\
    \midrule
    \multicolumn{2}{c}{(50000) Quaoar} \\
    \midrule
    Ephemeris version                & NIMAv16\tablefootmark{5}      \\
    Geocentric Distance              &  41.983157 au                 \\
    Apparent Magnitude               &  V = 18.8 \\ 
    Mass\tablefootmark{6}            & $(1.2 \pm 0.05) \times 10^{21}$~kg \\
    Rotation period\tablefootmark{7} & $17.6788 \pm 0.0004$~h \\
    \midrule
    \multicolumn{2}{c}{Weywot's orbital pole\tablefootmark{6}} \\
    \midrule
    RA  & 17$^h$44$^m~\pm$ 40$^m$ \\
    DEC & +50$^\circ~\pm$ 6$^\circ$ \\
    \midrule
    \end{tabular}
    \tablefoot{
    \tablefoottext{1}{The star position was taken from the \Gaia Data Release 3 (GDR3) star catalog \citep{GaiaColab2022} and is propagated to the event epoch using the formalism of \citet{Butkevich2014} applied with SORA \citep{GomesJr2022}.}
    \tablefoottext{2}{J, H, and K from the NOMAD catalog \citep{Zacharias2004}.}
    \tablefoottext{3}{Limb-darkened angular diameter estimated using a fit of the spectral energy distribution of the star with \citep{Castelli2003} atmosphere models, with the reddening definition of \citep{Fitzpatrick1999}.}
    \tablefoottext{4}{Renormalized unit weight error \citep{Lindegren2021}.}
    \tablefoottext{5}{Obtained from the Numerical Integration of the Motion of an Asteroid (NIMA, \citealt{Desmars2015}), based on astrometric positions derived from previous stellar occultations and the Minor Planet Center (MPC) database, available at \url{https://lesia.obspm.fr/lucky-star/obj.php?p=958}}
    \tablefoottext{6}{Obtained from the method of \cite{Vachier2012}.}
    \tablefoottext{7}{Assuming the double-peaked light curve from \cite{Ortiz2003}.}
    }
    \label{tab:star_obj}
\end{table}

Table \ref{tab:obs-circums} summarizes the observational circumstances. Among the stations, CFHT was equipped with the WIRCam \citep{WIRCam2004} at 2.15~$\mathrm{\mu}$m (Ks-band) with an 8.8 Hz acquisition rate. The nearby Gemini North telescope used the Alopeke instrument \citep{Alopeke2021} to simultaneously record the event with the z' (947 nm) and r' (620 nm) bands at a 10 Hz cadence. 
The images at the Tohoku University Haleakala Observatory (TUHO) in Hawaii and at the Transneptunian Automated Occultation Survey (TAOS II) in Baja California (Mexico) were taken with no filter to maximize the S/N. Q2R was not detected in these observations, likely due to insufficient S/N values. At the University of California, Santa Cruz (UCSC - California, US), the Dunrhomin Observatory (Colorado, US), Sommers-Bausch Observatory (Colorado, US), Nederland (Colorado, US), and a mobile station in Bonny Doon Eco Reserve (California, US), only the main body was recorded due to a low S/N.

%--------------------------------------------------------------------
\section{Data analysis}
\label{sec:data_analysis}

The data sets were analyzed using the standard photometry procedures of the Platform for Reduction of Astronomical Images Automatically (PRAIA, \citealt{MAssafin2010}). The flux of the occulted star was corrected by the flux of nearby reference stars to account for sky transparency variations and was normalized to the total star$+$Quaoar flux outside the occultation. 

The Stellar Occultation Reduction and Analysis package (SORA, \citealt{GomesJr2022}) was used to derive the ingress and egress instants of the occultation by the main body, assuming an opaque body without atmosphere. More details are presented in Appendix \ref{ap:main_body_fit}. The extremities of the resulting chords were finally used to fit an elliptical model to Quaoar's limb at the moment of the occultation.

In the case of rings, the events were fitted using the square box model described in Appendix~\ref{ap:model_rings} and pipelines based on the \textsc{SORA} package. This provides the radial width of the ring $W_{\rm r}$ (assuming a circular ring), its normal opacity, $p_{\rm N}$, and normal optical depth, $\tau_{\rm N}$, from which the  equivalent width, $E_{\rm p}= W_{\rm r} p_{\rm N}$, and equivalent depth, $A_{\tau}= W_{\rm r} \tau_{\rm N}$, are derived.

In the case of the CFHT and Gemini data, the acquisition rate and S/N were high enough to provide resolved profiles of the dense part of Q1R. 
Then, $E_{\rm p}$ and $A_{\tau}$ can be obtained directly from the numerical calculations of the integrals
$A_{\tau} = \int_{W_{\rm r}}{(v_{\rm r} \tau_{\rm N})dt}$ and
$E_{\rm p} = \int_{W_{\rm r}}{(v_{\rm r} p_{\rm N})dt}$
(i.e., without using the square model described above), $v_{\rm r}$ is the star velocity perpendicular to the ring, measured in the ring plane.

The dense part of Q1R has a radial profile reminiscent of the Saturnian F-ring core \citep{Bosh2002}. More precisely, it is consistent with a Lorentzian shape used here to provide the full width at half maximum (FWHM) of the profile and the event's timing, from which its location relative to Quaoar is obtained.

%--------------------------------------------------------------------
\section{Results}
\label{sec:results}

\begin{figure*}[!t]
\centering
\includegraphics[width=\hsize]{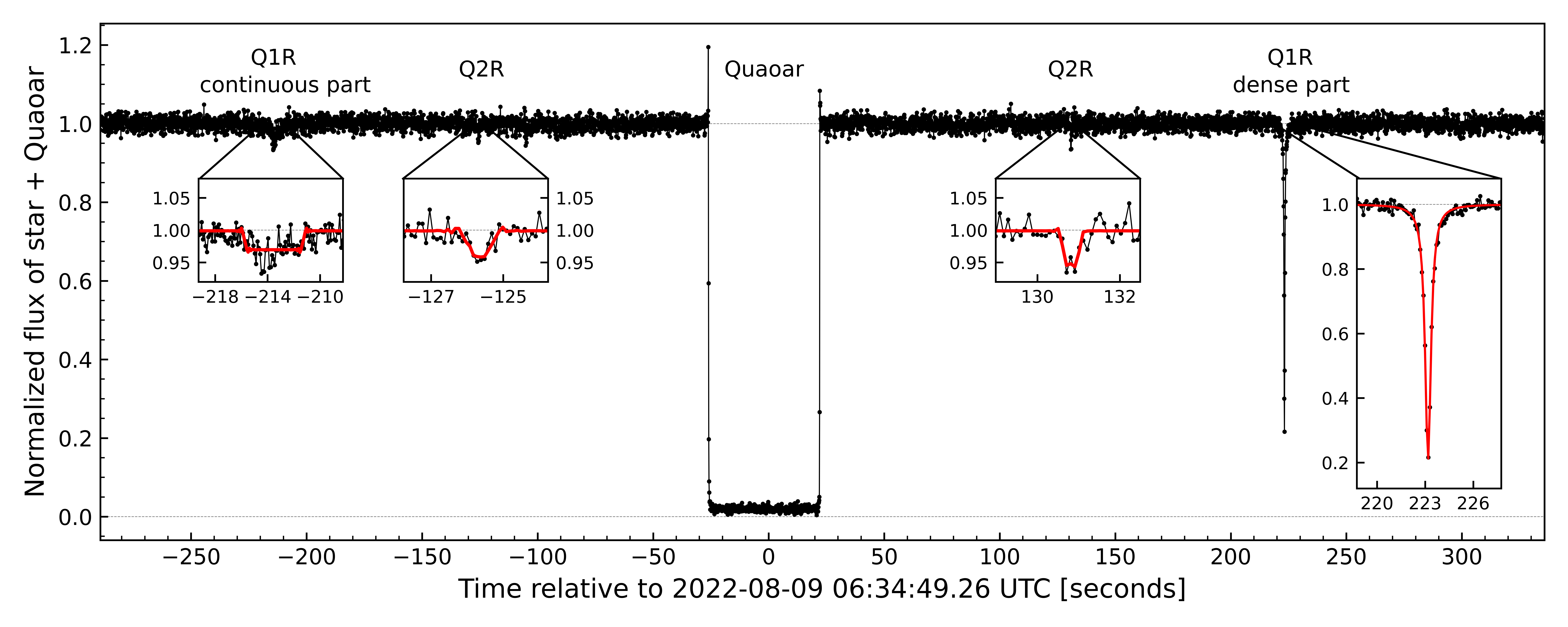}
\caption{%
Detection of Q1R, Q2R, and the main body in the Gemini (z') light curve. The normalized flux is plotted in black vs. time (UTC), while the best-fitting models are over-plotted in red. The shallow events caused by Q1R (ingress) and Q2R (ingress \& egress) are fitted by a square model. A Lorentzian function fits the dense part of Q1R at egress; see Sect. \ref{sec:data_analysis}. The spikes observed during the disappearance and the re-appearance of the star behind Quaoar stem from the diffraction by the sharp opaque limb of the body. Note: there is a variation in the flux close to -110 and 290 seconds, caused by sudden variations in the seeing, and they can be neglected. More detailed views of Q1R and Q2R obtained at other telescopes are displayed in Figs.~\ref{fig:Q2R_GNR}, \ref{fig:all_lcs}, \ref{fig:models_1}, and \ref{fig:models_2}.
}%
\label{fig:geminiR_Q1R_Q2R_body}
\end{figure*}

%--------------------------------------------------------------------
\subsection{Main body}
\label{sec:main_body}

The event on August 9, 2022  provided ten chords across Quaoar's main body. However, Gemini used a dual camera at two wavelengths, resulting in nine effective chords across the body, that is, $N$=18 chord extremities with the timings listed in Table \ref{tab:summary_instants}. 

We used the SORA package to fit an ellipse to those extremities projected in the sky plane. This ellipse is defined by $M=5$ parameters and the fit has $N-M= 13$ degrees of freedom. The fitted parameters are 
the ephemeris offsets $f_{\rm c}$ and $g_{\rm c}$ to apply to Quaoar's center in right ascension and declination, respectively, the apparent semi-major axis $a'$ of the limb,  its apparent oblateness $\epsilon'= (a'-b')/a'$, where $b'$ is the apparent semi-minor axis of the limb, and $P$ is the position angle of $b'$. The elliptical fit also considers a $\sigma_{\rm model}$, which is the uncertainty of the ellipse associated with the existence of putative topographic features on the surface of Quaoar. 
From the methodology presented by \citep{JOHNSON1973}, we estimated that Quaoar might support topographic features of about 5 km, given Quaoar's density $\rho = 1.99\,\rm{g}\,\rm{cm}^{-3}$ and strength $\rm{S} = 0.0303\times10^9\,dynes\,cm^{-2}$, consistent with an ice-rich composition. The standard deviation of the observed radial residuals is $\sim$8 km, providing an upper limit for topographic features consistent with this prediction. Considering the equivalent radius in area as 542 km, this corresponds to 1.5\%.
The fit results are listed in Table~\ref{tab:limb_fit}. The chords and the best-fitting elliptical limb are plotted in Fig.~\ref{fig:rings_sky}, with more details given in Fig.~\ref{fig:main_body}.

\begin{table}[ht]
\centering
\small
\caption{Retrieved parameters of Quaoar's body and its rings\tablefootmark{1}.}
\begin{tabular}{l l} 
\toprule \toprule
\multicolumn{2}{c}{Quaoar's limb fitting} \\
\midrule
Apparent semi-major axis        & $a' = 579.5 \pm 4.0$~km          \\ 
Apparent oblateness             & $\epsilon'= 0.12 \pm 0.01$     \\    
Position angle                  & $P= 345.2 \pm 1.2$~degrees          \\
Ephemeris offset                & $f_{\rm c}= -24.7 \pm 2.0$~km    \\
                                & $g_{\rm c}= -13.9 \pm 2.1$~km    \\
$R_{\rm equiv}$                 & $543 \pm 2$ km \\
$\chi^2$ per degree of freedom  & 0.986 \\
\midrule
Quaoar's geocentric             & $\alpha$= 18$^h$21$^{m}$42$^s$.8677703 $\pm$ 0.249 mas \\
\hspace{5mm}position (ICRS)     & $\delta$=  -15$^\circ$12$'$ 45$''$.829691 $\pm$ 0.230 mas \\
\midrule
Q1R's preferred pole              &  $\alpha_{\rm P}$= 17$^h$19$^{m}$16$^s$.8 $\pm$ 55$^s$.2 \\
\hspace{5mm}position (ICRS)\tablefootmark{2}  &  $\delta_{\rm P}$= +53$^\circ$27'$\pm$ 18' \\ 
\midrule 
\multicolumn{2}{c}{Q1R's geometry} \\
\midrule
Radius                           & $4,057 \pm 6$ km    \\
Opening angle                    & $B= -20.0 \pm 0.3$ degrees \\ 
Position angle                   & $P= 350.2 \pm 0.2$ degrees \\ 
$\chi^2$ per degree of freedom   & 1.633 \\
\midrule
\multicolumn{2}{c}{Q2R's geometry\tablefootmark{3} } \\
\midrule
Radius                          & $2,520 \pm 20$ km \\ 
\midrule
\end{tabular}
\tablefoot{
\tablefoottext{1}
{Quaoar's limb, Quaoar's geocentric position, and Q1R's geometry are given at epoch August 9$^{\rm {th}}$, 2022, 06:34:03.560 UTC. The fits use the star position and the NIMAv16 ephemeris for Quaoar (Table~\ref{tab:star_obj}). }
\tablefoottext{2}
{The preferred solution is co-planar with Quaoar's equatorial plane and Weywot's orbit within $5 \pm 7$ deg. The complementary position $\alpha_{\rm P}$= 10$^h$00$^{m}\pm05^m$ and $\delta_{\rm P}$= +79$^\circ$21$'\pm08'$, is not considered here as it provides a less satisfactory fit to the Q1R detections and yields an inclination of 45 degrees with respect to Weywot's orbit, see Sec. \ref{sec:Q1R}}
\tablefoottext{3}
{Assuming a circular ring co-planar with Q1R. The error bar for the Q2R ring radius may be underestimated since we only have two effective detections, and several parameters were assumed.}
}
\label{tab:limb_fit}
\end{table}

%--------------------------------------------------------------------
\subsection{Ring Q1R}
\label{sec:Q1R}
The Q1R ring was detected on both sides of the main body in the Gemini and CFHT light curves, and their physical properties were determined as described in Sect. \ref{sec:data_analysis}. Although the detections of the dense part of the Q1R in the Gemini and CFHT light curves indicate the presence of diffuse material around the ring, the quality of the TAOS II and TUHO light curves only allows for the narrow and dense part of the ring to be detected. Therefore, the parameters $E_{\rm p}$ and $A_{\tau}$ were determined by the square box model for these data.

Due to the low optical depth of the Q1R segment intercepted before the closest approach, only the light curves obtained at Gemini and CFHT have sufficient S/N for detection (see an example in Fig.~\ref{fig:geminiR_Q1R_Q2R_body} and more details in
Figs.~\ref{fig:Q2R_GNR}, \ref{fig:all_lcs}, \ref{fig:models_1}, and \ref{fig:models_2}).
The Gemini detection in the r' bandpass seems more sharply defined than its counterpart in z', but this effect remains marginal considering the S/N. The CFHT light curve shows a drop simultaneously, but it may be affected by a seeing deterioration that compromises an accurate determination of the ring boundaries. Even so, the central times of the Gemini and CFHT detections are all consistent at the 1-$\sigma$ level. 
The parameters of Q1R for all the detections are presented in Table~\ref{tab:rings_local}.

We have combined the data from this work and the previous observations to improve the Q1R orbital parameters, assuming a fixed ring pole orientation between 2018 and 2022, as per \cite{Morgado2023}. We tested a range of pole orientations and ring radii using a $\chi^2$ statistic, resulting in two complementary solutions presented in Table~\ref{tab:limb_fit}. The preferred solution has a $\chi^2$ per degree of freedom $\chi^2_{\rm pdf} = 1.6$ and shows a better agreement with observations than the mirror solution with $\chi^2_{\rm pdf}= 4.1$. Moreover, the preferred solution is co-planar with Weywot's orbit to within $5 \pm 7$~deg, as expected from a primordial disk surrounding Quaoar that evolved into a ring and formed its satellite, while the mirror solution is inclined by $45 \pm 7$~degrees with respect to 
Weywot's orbit.

The position angles of Quaoar's projected pole ($345.2 \pm 1.2$~deg) and Q1R ($350.2 \pm 0.2$~deg) are misaligned by $\sim$5~degrees (Table~\ref{tab:limb_fit}). Thus, our results are consistent with Q1R orbiting close to Quaoar's equatorial plane, assuming that the body is an oblate spheroid. Part of this misalignment could stem from the fact that Quaoar is a triaxial ellipsoid (or a body with a more complex shape). Hence, the position angle of Quaoar's limb does not necessarily coincide with the position angle of Quaoar's pole.

%%%%%%%%%%%%%%%%%%%%%%%%%%%%%%%%%%%%  Q2R
\subsection{The discovery of a new ring around Quaoar}
\label{sec:Q2R}

\begin{figure}[ht]
\centering
\includegraphics[width=\hsize]{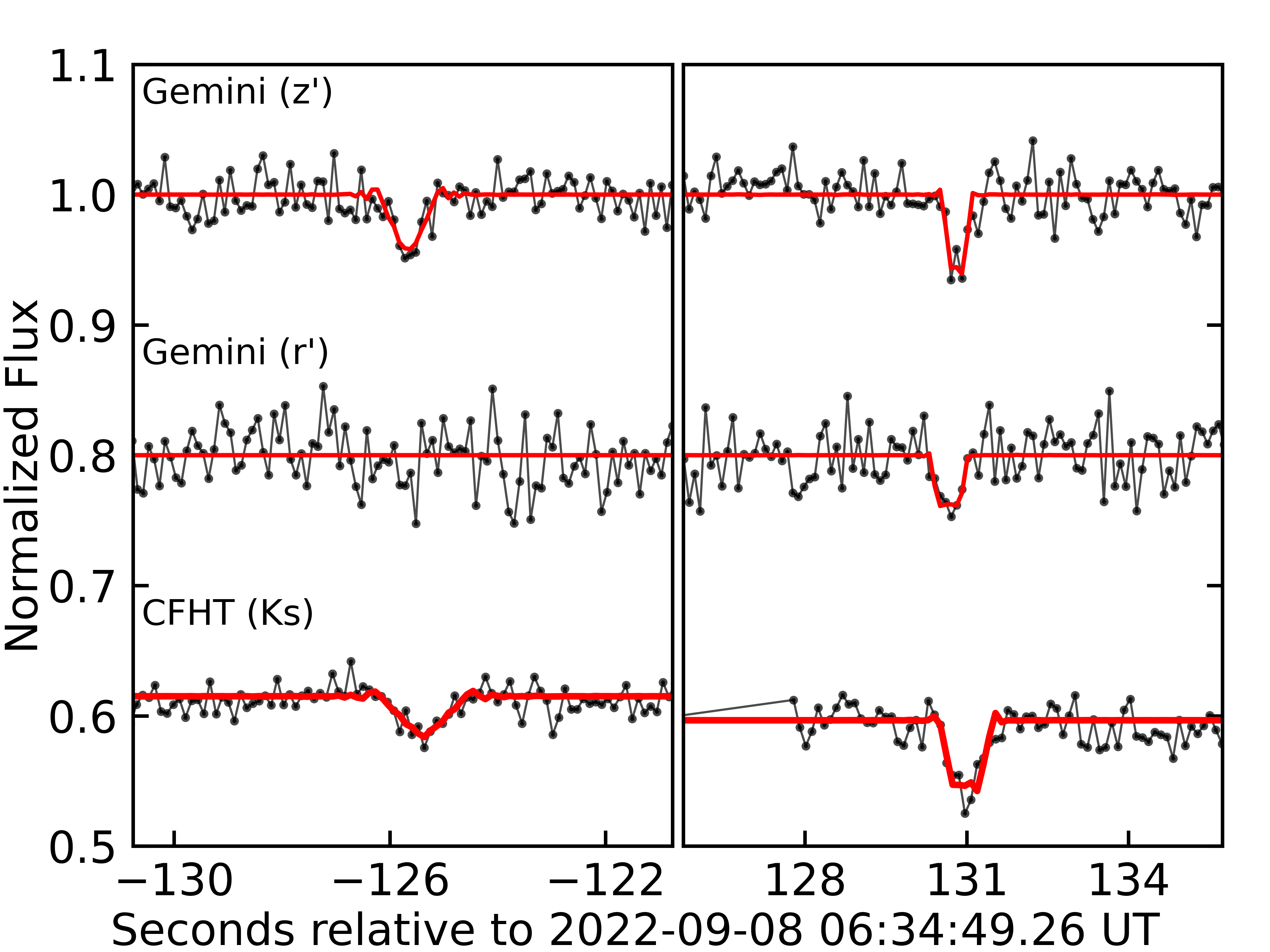}
\caption{
Detection of the Q2R ring in the Gemini and CFHT light curves. The data points are plotted in black, and the red lines are the square box model fits derived from the quantities listed in Table~\ref{tab:rings_local}. An arbitrary offset was applied in the vertical direction for clarity. The time axis is relative to August 9, 2022 at 06:34:49.26 UTC, the time of the closest approach of Mauna Kea to Quaoar's shadow center.
}
\label{fig:Q2R_GNR}
\end{figure}

The unique photometric quality of the Gemini and CFHT data allowed for the detection of additional material around Quaoar (Fig.~\ref{fig:geminiR_Q1R_Q2R_body}).  
These data sets reveal additional secondary events symmetrically located with respect to Quaoar (Fig.~\ref{fig:Q2R_GNR}). Two such events are simultaneously detected in the Gemini z'-band and CFHT Ks-band light curves before the closest approach, with detections reaching around 5.5$\sigma$ and 5.2$\sigma$, respectively. Conversely, the light curves display simultaneous events after the closest approach with significant detections, standing at 5.7$\sigma$, 3.7$\sigma$, and 4.7$\sigma$ for the Gemini z', Gemini r', and CFHT data sets, respectively. Assuming that the light curves have a normal distribution, the probability that individual points of equivalent width E$_{\rm{p}}(3\sigma) > 12$ km occur randomly in each light curve is $\rm {p} \approx 1.4\times 10^{-3}$, with p approaching zero for values larger than E$_{\rm{p}}$.   While the two light curves of the Gemini instrument may be correlated as they were taken at the same telescope, the Gemini and CFHT data are independent in terms of fast-seeing fluctuations. Using Poisson statistics, we estimate that the probability that the simultaneous events in the Gemini and CFHT data occur randomly due to the seeing fluctuations is very low, with $\rm {p} \approx 10^{-6}$.

Some dips in flux were observed in the vicinity of Q2R detection in the region after the closest approach (Fig. 2). These detections are at the detection limit of our data, so we cannot say whether these are due to seeing variations or additional structures. Furthermore, we do not observe counterparts of these dips in the region before the closest approach.

All these detections are consistent with a new circular ring (Q2R) co-planar and concentric with Q1R, orbiting at $2,520 \pm 20$~km from Quaoar (Table~\ref{tab:limb_fit}). The radial width, optical depth, equivalent width, and equivalent depth of Q2R are listed in Table~\ref{tab:rings_local}. They were obtained as described in Appendix~\ref{ap:model_rings}. 

\section{Discussion and conclusions}
\label{sec:Conclusion}

\begin{figure}[!t]
\centering
\includegraphics[width=\hsize]{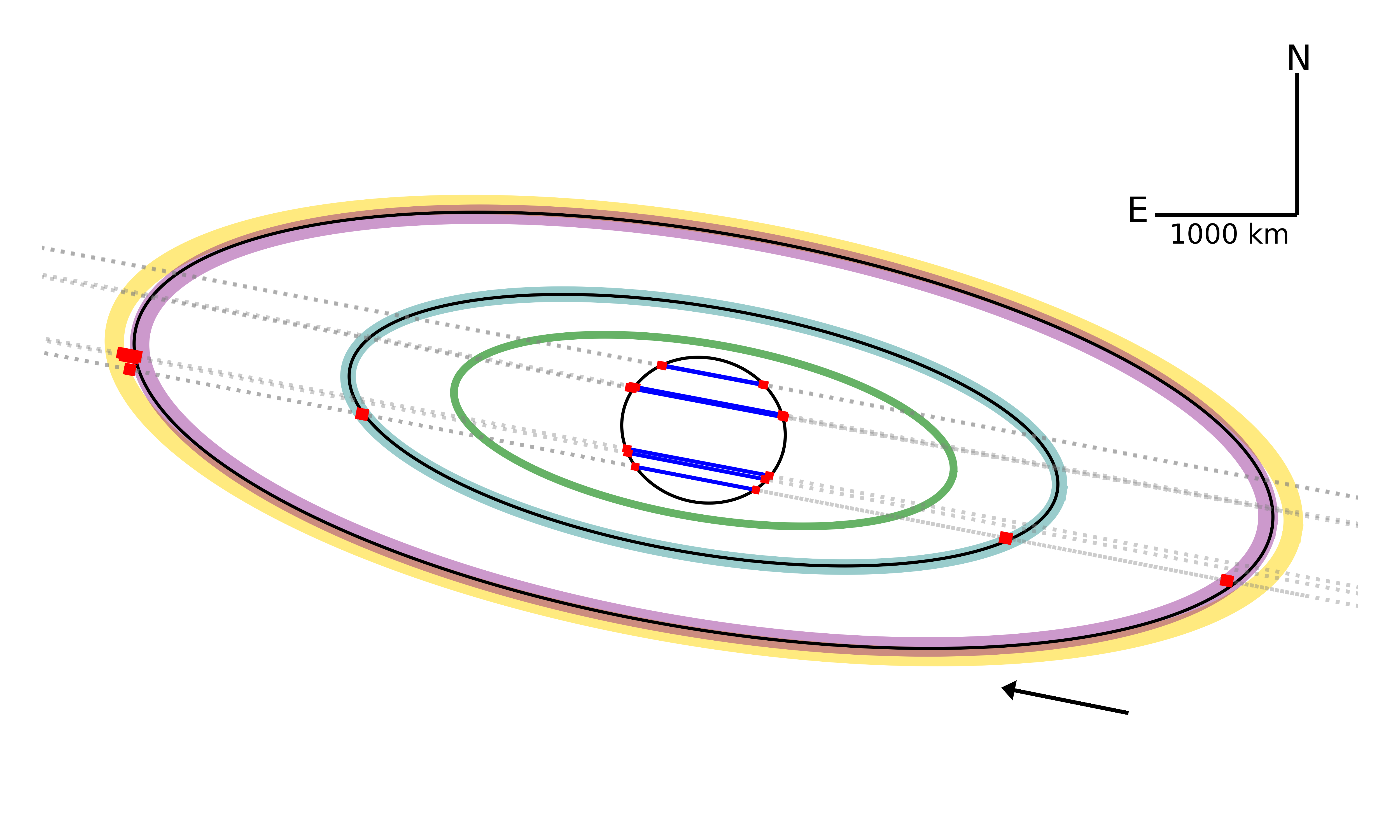}
\caption{
Representation of our results on Quaoar's shape (center) and the detection of the two rings Q1R (outer ring) and Q2R (inner ring). The red segments correspond to the 1-$\sigma$ error bars on the particular events. The orbit of Q1R is determined from a simultaneous fit using the present work and previous detections of 2018, 2019, 2020, and 2021 reported by \cite{Morgado2023} (see Section~\ref{sec:Q1R}). The solution for the orbit of the new Q2R ring assumes that this ring is co-planar and concentric with Q1R. The central part of the plot (occultation by the solid body) is enlarged in Fig.~\ref{fig:main_body}. 
In yellow, we show the 1/3 SOR resonance with Quaoar, and in teal is the 5/7 SOR resonance. The purple ellipse represents 6/1 MMR with Weywot (considering the double-peaked rotation period), and the green ellipse presents the expected Roche limit, considering particles with a bulk density of $\rho = 0.4\,\rm{g}\,\rm{cm}^{-3}$. The arrow shows the star's motion relative to Quaoar. Note: the orbital radius of Weywot is about three times larger than that of Q1R, and thus it is not shown in this representation. 
}
\label{fig:rings_sky}
\end{figure}

The stellar occultation of August 9$^{\rm {}}$, 2022 provided nine effective occultation chords obtained in Hawaii, Mexico, and the continental United States. They constrained Quaoar's shape, providing an apparent semi-major axis of $a'= 579.5 \pm 4.0$~km, an apparent oblateness $\epsilon'= 0.12 \pm 0.01$, and an area-equivalent radius of $R_{\rm equiv} = a' \sqrt{1-\epsilon'}= 543 \pm 2$~km. Using an absolute magnitude of $H_{\rm V} = 2.73 \pm 0.06$ \citep{Fornasier2013}, this yields a geometric albedo of $p_{\rm V} = 0.124 \pm 0.006$. \bla
Our value of $R_{\rm equiv}$ differs from that of \cite{BragaRibas2013}, $555.0 \pm 2.5$~km, by about 12~km, that is, at the 4-$\sigma$ level. This difference could be \bla evidence that Quaoar is not an oblate spheroid, but rather a triaxial ellipsoid or a body with a more complex shape; alternatively, it may be caused by a change in the aspect-angle since 2011, as observed from Earth. 

The continuous region and the dense part of Q1R were detected during this event. The dense part was radially resolved and showed a Lorentzian profile reminiscent of Saturn's F ring \citep{Bosh2002} or Neptune's ring arcs \citep{Nicholson90, Sicardy1991}. This contrasts with the sharp edges observed for Chariklo's main ring C1R \citep{Berard2017, Morgado2021}. 
The dense part of Q1R is detected over a radial width of $\sim$60~km with a peak optical depth of $\tau_{\rm N} \approx 0.4$, an FWHM Lorentzian width of 5~km and an equivalent width $E_{\rm p}$ of around 2~km (see Table~\ref{tab:rings_local}). 
The values presented here are more precise than (but consistent with) those published by \cite{Morgado2023}.

The detections of the dense part of Q1R in 2021 imply a minimum arc length of 365~km, corresponding to an azimuthal extension of about 5.1~degrees \citep{Morgado2023}. The detections from CFHT, Gemini, TUHO, and TAOS II in 2022 (Figs.~\ref{fig:models_1}, \ref{fig:models_2}) suggest a minimum arc-length of 226 km or $\sim$3.2~degrees. 
Since 2011, we have obtained 19 cuts of the Q1R with a sufficient S/N to detect the densest region. The limited azimuthal extent of the two detections of the dense regions means that they could both be parts of the same arc-like structure. If this structure was detected two times among the 19 cuts, then its extent can be estimated using Poisson statistics. This analysis yields an arc length with a 70\% chance of falling between 18 and 72 degrees. In this case, all of the detections in either 2021 or 2022 would each be correlated samples of one part of this arc.

The more tenuous component of Q1R is radially resolved and shows no marked structures, being consistent with a square model within our S/N limits. The best light curves (Gemini and CFHT) provide a typical width of 80-100~km and a typical normal optical depth of 0.003 (and, thus, an equivalent width of $\sim$0.3~km) for that component at the longitude where it was detected (Table~\ref{tab:rings_local}).

Our detections, combined with those reported by \cite{Morgado2023}, improved the orbital elements of Q1R. They are consistent with a circular ring of radius $4,057.2 \pm 5.8$~km (Table~\ref{tab:limb_fit}) corresponding to about 7.5$\times$ Quaoar's area-equivalent radius $R_{\rm equiv}$. 
This value coincides with the 6/1 MMR with Weywot ($4,020 \pm 60$~km) and within 3-$\sigma$ of the 1/3 SOR with Quaoar ($4,200 \pm 60$~km), considering the double-peaked rotation period of $17.6788 \pm 0.0004$ h \citep{Ortiz2003}. 

Our preferred solution for Q1R's orbital pole is consistent with this ring being co-planar with Weywot's orbit.
Moreover, the apparent semi-major axes of Quaoar's limb and Q1R's orbit are aligned at the 4-$\sigma$ level, suggesting that Q1R lies in Quaoar's equatorial plane, as expected from a colliding ring system \citep{Kokubo2000}.

Our data reveal a new ring (Q2R) around Quaoar. The detections are consistent with a circular ring of radius $2,520 \pm 20$~km co-planar with Q1R  (Table~\ref{tab:limb_fit}, Figs.~\ref{fig:Q2R_GNR} and \ref{fig:rings_sky}). It has a typical width of 10~km, an optical depth of about 0.004, and an equivalent depth of about 0.04~km (see accurate values in Table~\ref{tab:rings_local}). Although closer to Quaoar than Q1R, Q2R is at 4.6$\times$ Quaoar's radius, also outside Quaoar's Roche limit, which is estimated to be around 1,780~km, assuming the ring particle density as $\rho = 0.4\,\rm{g}\,\rm{cm}^{-3}$ \citep{Morgado2023}.

Using previously determined values for Quaoar's mass and rotation period (Table~\ref{tab:star_obj}), we derived a 5/7 SOR radius of  $2,525\pm35$~km. This coincides with Q2R's radius, $2,520 \pm 20$~km, to within the  1-$\sigma$ error bars. Like the 1/3 SOR, the 5/7 SOR is a second-order resonance and, as such, may play an essential role in the confinement of Q2R. However, more solid determinations of Q2R's orbit and Quaoar's shape are required. 

Table~\ref{tab:rings_local} shows that differences in the equivalent widths of Q1R and Q2R are observed between the z', r', and Ks filters. The significance and interpretation of these differences require more analysis and will be discussed in a forthcoming publication. Moreover, the high S/N obtained at Gemini and CFHT will be used to detect or put a stringent upper limit on a putative atmosphere. Finally, the comparison of past (and possibly future) results derived from multi-chord occultations by Quaoar's main body will be used to constrain its shape better. This will be important for better understanding the dynamics of Quaoar’s ring system, particularly under the effect of spin-orbit resonances with the body.

%--------------------------------------------------------------------
\begin{acknowledgements}
  C.L.P is thankful for the support of the CAPES and FAPERJ/DSC-10 (E26/204.141/2022). 
  This work was carried out within the “Lucky Star" umbrella that agglomerates the efforts of the Paris, Granada, and Rio teams, funded by the European Research Council under the European Community’s H2020 (ERC Grant Agreement No. 669416). 
  This study was financed in part by the National Institute of Science and Technology of the e-Universe project (INCT do e-Universo, CNPq grant 465376/2014-2). 
  This study was financed in part by CAPES - Finance Code 001. 
  The following authors acknowledge the respective CNPq grants: B.E.M. 150612/2020-6; F.B.R. 314772/2020-0; R.V.M. 307368/2021-1; M.A. 427700/2018-3, 310683/2017-3, 473002/2013-2; J.I.B.C. 308150/2016-3, 305917/2019-6.
  R.C.B acknowledge the FAPERJ grant E26/202.125/2020.
  E.F.-V. acknowledges financial support from the Florida Space Institute and the Space Research Initiative.
  J.L.O., P.S-S., M.V-L, and M. K. acknowledge financial support from the grant CEX2021-001131-S funded by MCIN/AEI/ 10.13039/501100011033, they also acknowledge the financial support by the Spanish grants PID2020-112789GB-I00 from AEI and Proyecto de Excelencia de la Junta de Andalucía PY20-01309.
  Funding for RECON was provided by grants from USA: NSF AST-1413287, AST-1413072, AST-1848621, and AST-1212159. We thank RECON observers Doug Thompson, Ken Conway, Dorey Conway, Terry Miller, David Schulz, Michael von Schalscha, and Matt Christensen for their efforts in collecting data.
  Based on observations obtained with WIRCam, a joint project of CFHT, Taiwan, Korea, Canada, France, at the Canada-France-Hawaii Telescope (CFHT)  which is operated from the summit of Maunakea by the National Research Council of Canada, the Institut National des Sciences de l'Univers of the Centre National de la Recherche Scientifique of France, and the University of Hawaii. The observations at the Canada-France-Hawaii Telescope were performed with care and respect from the summit of Maunakea which is a significant cultural and historic site.
  We thank Marc Baril and Tom Vermeulen for their time dedicated to the observation performed at Canada-France-Hawaii Telescope (CFHT). 
  Based on observations obtained at the international Gemini Observatory, a program of NSF’s NOIRLab, which is managed by the Association of Universities for Research in Astronomy (AURA) under a cooperative agreement with the National Science Foundation on behalf of the Gemini Observatory partnership: the National Science Foundation (United States), National Research Council (Canada), Agencia Nacional de Investigaci\'{o}n y Desarrollo (Chile), Ministerio de Ciencia, Tecnolog\'{i}a e Innovaci\'{o}n (Argentina), Minist\'{e}rio da Ci\^{e}ncia, Tecnologia, Inova\c{c}\~{o}es e Comunica\c{c}\~{o}es (Brazil), and Korea Astronomy and Space Science Institute (Republic of Korea). 
  This work made use of data from GN-2022B-DD-101 observing program and were obtained with the High-Resolution Imaging instrument(s) ‘Alopeke (and/or Zorro). ‘Alopeke (and/or Zorro) was funded by the NASA Exoplanet Exploration Program and built at the NASA Ames Research Center by Steve B. Howell, Nic Scott, Elliott P. Horch, and Emmett Quigley. ‘Alopeke (and/or Zorro) was mounted on the Gemini North (and/or South) telescope of the international Gemini Observatory, a program of NSF’s NOIRLab, which is managed by the Association of Universities for Research in Astronomy (AURA) under a cooperative agreement with the National Science Foundation. 
  This work was enabled by observations made from the Gemini North telescope, located within the Maunakea Science Reserve and adjacent to the summit of Maunakea. We are grateful for the privilege of observing the Universe from a place that is unique in both its astronomical quality and its cultural significance.
  This research used \textsc{sora}, a python package for stellar occultations reduction and analysis, developed with the support of ERC Lucky Star and LIneA/Brazil, within the collaboration of Rio-Paris-Granada teams.
  This work profited from unpublished occultations by Quaoar made at SOAR (SO2019A-003) and the Pico dos Dias Observatory (OP2019A-004) to improve the accuracy of the ephemeris~NIMAv16.
\end{acknowledgements}


\bibliographystyle{aa}
\bibliography{references.bib}

\begin{appendix} 
%%%%%%%%% Appendix C
\section{Main body occultation and elliptical fit}
\label{ap:main_body_fit}
The light curves obtained with the Gemini and CFHT telescopes present prominent diffraction signatures in the main body occultation mainly due to the high acquisition rate, the large wavelength, and high S/N of the light curves. During the occultation, Quaoar was at a geocentric distance of 41.983157 au, implying a Fresnel scale, $F = \sqrt{\lambda D / 2}$, from 1.5 to 2.6 km, for visible to near-infrared (CFHT Ks-band) wavelengths, respectively. Considering the apparent velocity of the event as 17.57 km s$^{-1}$ and the acquisition rate of the Gemini and CFHT instruments, we obtained a spatial resolution of these light curves of $\sim$1.8 km. These high-S/N light curves have a Fresnel diffraction effect on the same order of magnitude as the instrumental response (e.g., the integration time). 
\begin{figure}[!h]
\centering
\includegraphics[width=\hsize]{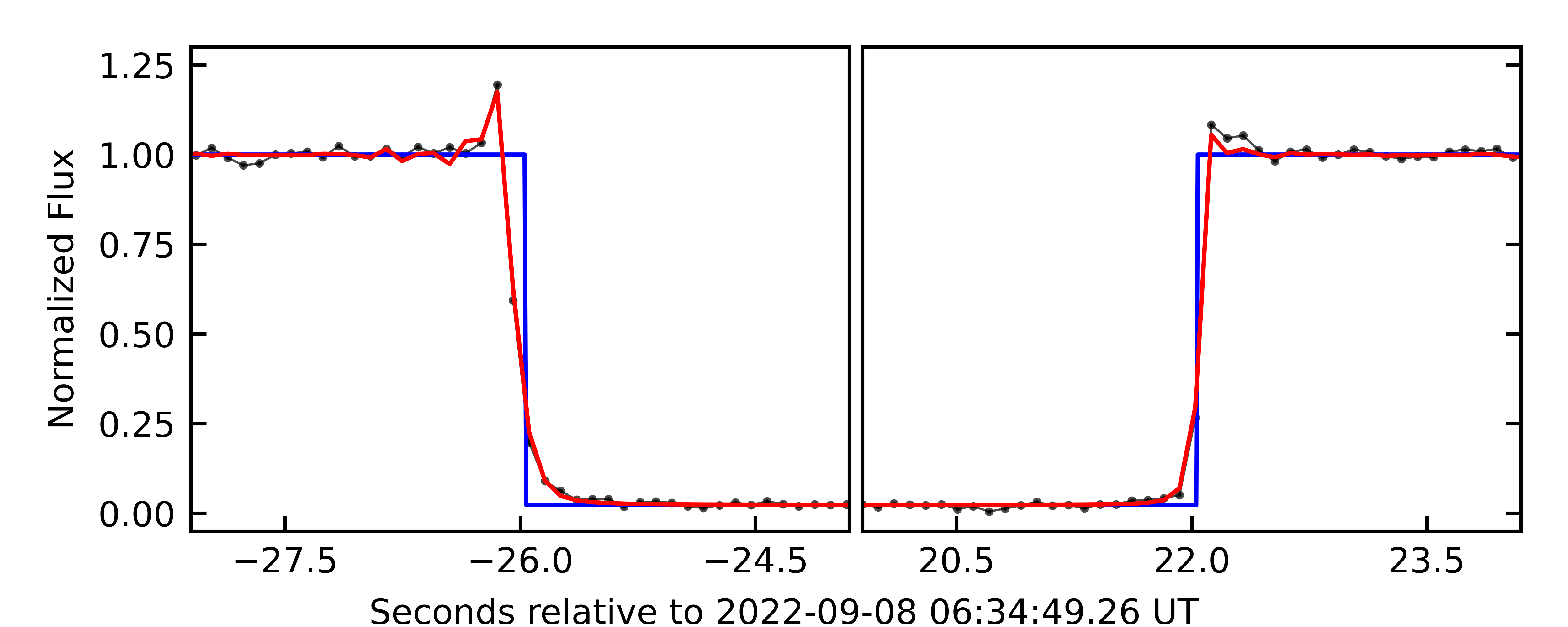}
\caption{Gemini z' light curve (black dots) and the modeled light curve (red) considering the local normal star velocity, the wavelength and apparent star size. The geometric model is in blue. The normalized flux is plotted as a function of the seconds before and after the local closest approach (2022-08-09 06:34:49.26 UT).}
\label{fig:quaoar_model}
\end{figure}

\begin{table}[!h]
\centering
\small
\caption{
Ingress and egress times for Quaoar's main body. 
}
\begin{tabular}{l c c} 
\toprule \toprule
Site                    &   Ingress              &   Egress               \\ 
                        &   (hh:mm:ss.s)         & (hh:mm:ss.s)         \\
\midrule
CFHT                    & 06:34:23.34 (0.02) & 06:35:11.42 (0.01) \\
Gemini (z')             & 06:34:23.29 (0.01) & 06:35:11.29 (0.01) \\    
Gemini (r')             & 06:34:23.27 (0.03) & 06:35:11.28 (0.02) \\
TUHO                    & 06:34:23.5 (0.2)   & 06:35:18.1 (0.2)   \\
TAOS II                 & 06:30:38.3 (0.2)   & 06:31:35.1 (0.1)   \\
Dunrhomin               & 06:30:03.3 (0.4)   & 06:31:02.8 (0.9)   \\
Sommers-Bausch          & 06:30:02.6 (0.2)   & 06:31:03.8 (0.5)   \\
UCSC                    & 06:31:20.3 (0.1)   & 06:32:00.9 (0.1)   \\
Nederland               & 06:30:03.4 (0.5)   & 06:31:04.4 (0.8)   \\
Bonny Doon              & 06:31:20.7 (0.4)   & 06:32:01.3 (0.4)   \\
\midrule
\end{tabular}
\tablefoot{Times are in UT and the error bars in parentheses are in seconds and given at the 1-$\sigma$ level. The significant differences in the CFHT and Gemini instants are due to the length of the chords, with the CFHT being $\sim1$ km longer than the Gemini, as expected from the limb.}
\label{tab:summary_instants}
\end{table}

With the limb-darkened angular diameter of the star estimated and considering the observation band pass and its FWHM, we also fit the local normal velocity ($v_{\perp}$). By minimizing the $\chi^2$ statistic, we found the velocity and the ingress (egress) time that best explained the observed light curve. This was applied to the Gemini z'-band data since this was our best light curve. The radial velocities obtained are $v_{\perp}$ = 10.2 km s$^{-1}$ and $v_{\perp}$ = 16.6 km s$^{-1}$ for ingress and egress instants, respectively. The separation between the chords obtained from the Gemini and CFHT when projected onto the sky plane was about 160 meters, a value smaller than the star's apparent diameter. Therefore, the velocities were considered the same at both sites. This procedure resulted in an optimal fit for both diffraction spikes and the baseline occultation by the main body. Figure \ref{fig:quaoar_model} presents an example of this fit with the Gemini z' light curve plotted with the modeled curve.

\begin{figure}[!h]
\centering
\includegraphics[width=\hsize]{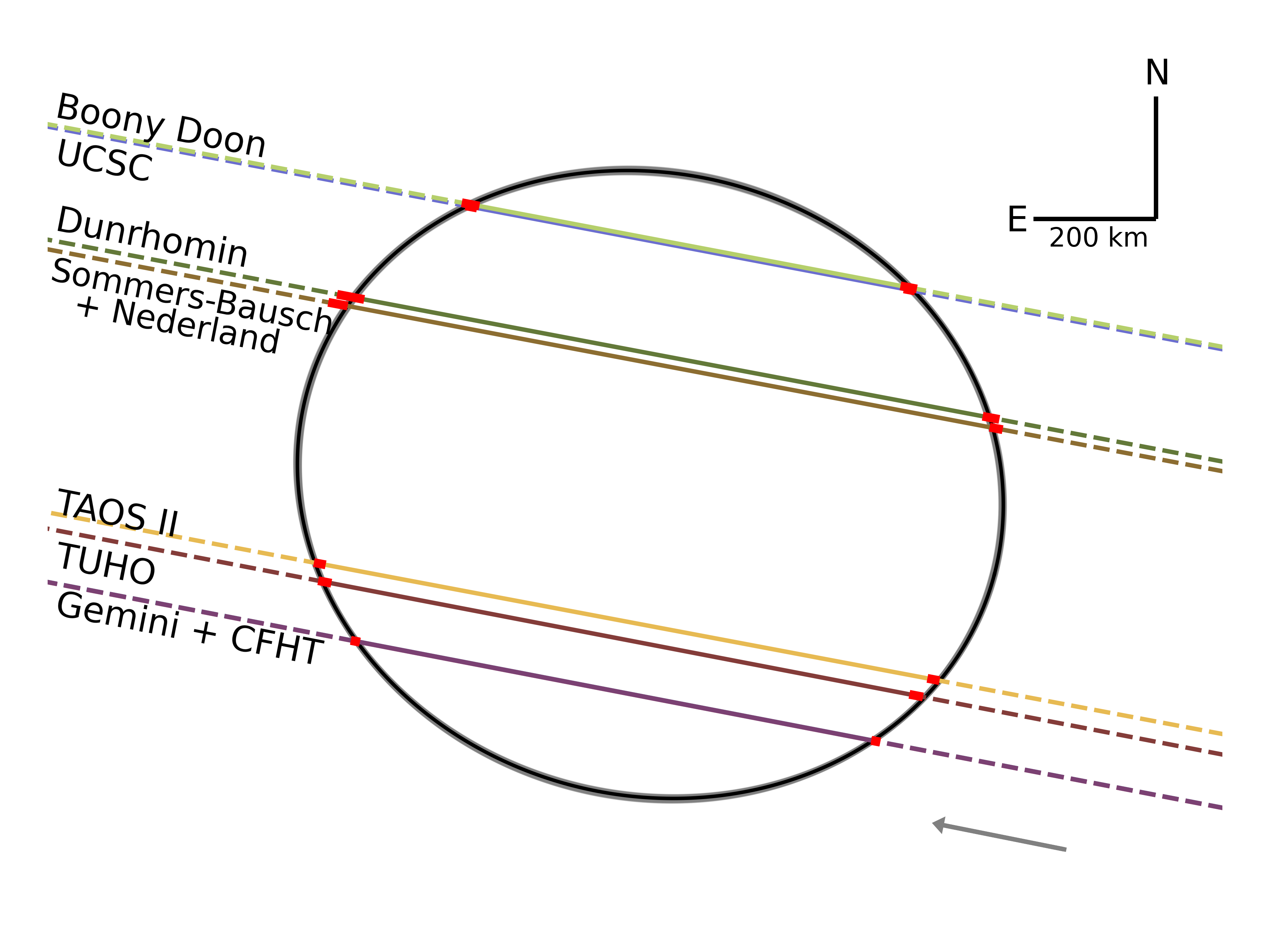}
\caption{%
Close-up view of Fig.~\ref{fig:rings_sky} shows the  best elliptical fit (black ellipse) and ellipses within 1-$\sigma$ region (grey) to the ten chords derived from the timings in Table~\ref{tab:summary_instants}. The grey 1-$\sigma$ fits differ only slightly from the black best-fit. The red segments represent the 1-$\sigma$ error bars in the ingress and egress times. At this scale, the Gemini and CFHT chords are superimposed, and their error bars would be too small to be seen (i.e., they are here represented by red squares, much bigger than their actual sizes).
}%
\label{fig:main_body}
\end{figure}

Ingress and egress times for the main body occultation are presented in Table \ref{tab:summary_instants}. The instant uncertainties of the Gemini and CFHT light curves are limited by occultation modeling, not data quality. The elliptical fit parameters to the chord extremities are presented in Fig. \ref{fig:main_body}.

\clearpage

%%%%%%%%%% Appendix A
\section{Observational circumstances}
The observational circumstances for this occultation campaign are presented in Table \ref{tab:obs-circums}. We present in this work the analysis of ten data sets with positive detections for the main body, of which four are detections of the Q1R ring and three are detections of both the Q1R and Q2R rings. In addition, we have more eighteen sites that participated in the observation campaign, but reported cloudy skies. 
The target star was too faint for the setup in  Woodside (CA, US) and Summerland BC, CA, even with 4 and 3 seconds of exposure, respectively; in addition, the Summerland data was acquired while clouds were drifting through the field.

\begin{table*}[!ht]
\centering
\caption{Observational circumstances.}
\begin{tabular}{c c c c c c} \toprule \toprule
            & \textbf{Longitude [$^\circ$ ' '']}   &                    & \textbf{Aperture (mm)}      & \textbf{Exp. Time} &  \textbf{Status}        \\
\textbf{Site}   & \textbf{Latitude [$^\circ$ ' '']}    & \textbf{Observers} & \textbf{Detector}           & \textbf{Cycle}     & \textbf{Light curve}    \\
            &\textbf{ Altitude [m]}           &                    & \textbf{Filter}             & \textbf{(s)}       & \textbf{Dispersion}    \\ 
\midrule
CFHT                      & -155 28 07.93    &  H. Januszewski        &  3,580                & 0.113             &  Positive   \\
MaunaKea                  & 19 49 31.01      &  M. Baril, B. Epinat   & WIRCam                & 0.113             &   0.013     \\
HI, US                    & 4,206.1           &  T. Vermeulen          & Ks                   &                  &             \\                   
\midrule
Gemini-North              & -155 28 08.6       &    T. Seccull          & 8,100                 & 0.100             &  Positive   \\
MaunaKea                  & 19 49 26           &   A. Stephens          & 'Alopeke              & 0.101             &  z': 0.012  \\
HI, US                    & 4,213              &                        & r' \& z'              &                   &  r': 0.022  \\  
\midrule
TUHO                      & -156 15 27       &  E. Tatsumi            & 600                   & 1.500             &  Positive   \\
Haleakala                 & 20 42 25.14      &  F. Yoshida            & ASI178MM              & 1.634             &    0.098    \\
HI, US                    & 3,051.8           &  M. Kagitani           & None                 &                   &             \\      
\midrule
TAOS II                 & -115 27 47.88    &  J. H. Castro-Chacón      & 1,300                       & 0.200             &  Positive   \\
San Pedro Martir          & 31 02 36.26      &  S. Sánchez-Sanjuán      & Andor Ixon 888              & 0.205             &   0.274     \\
BC, MX                    & 2,824.8           &                  & None                       &                  &              \\
\midrule
Dunrhomin Obs.            & -105 09 46.57    &                  & 280                        & 2                 &  Positive   \\
Longmont                  & 40 15 08.46      &  M. Buie         & QHY174M-GPS                & 2                 &     0.229   \\
CO, US                    & 1,592.8           &                  & None                      &                   &              \\
\midrule
Sommers-Bausch Obs.       & -105 15 47.09    &  J. Keller       &  508                       & 0.9               &  Positive   \\
Boulder                   & 40 00 13.32      &  J. Johnston     & QHY174M-GPS                & 0.9               &   0.259     \\
CO, US                    & 1,645           &                  & None                        &                   &             \\
\midrule
                      & -105 26 44.05    &   M. Skrutskie         & 200                  & 2                 &  Positive   \\
Nederland                 & 39 59 13.95      &   A. Verbiscer         & QHY174M-GPS          & 2                 &  0.313      \\
CO, US                    & 2,492.6           &                        & 570 nm              &                   &             \\ 
\midrule
Bonny Doon Eco Reserve    & -122 08 18.62    &                  &  203                       & 0.260             &  Positive   \\
Santa Cruz                & 37 03 03.12      &  K. Bender       & Watec 910HX/RC             & 0.260             &     0.538   \\
CA, US                    & 488.4            &                  & None                       &                   &             \\
\midrule
UCSC                      & -122 04 46.20    &                  &  203                       & 0.500             &  Positive   \\
Santa Cruz                & 37 01 03.47      &  R. Nolthenius   & Watec 910HX/RC             & 0.534             &  0.556      \\
CA, US                    & 344.0            &                  & None                       &                   &             \\
\midrule
                      & -119 40 12       &                        &  457                &  3                  &             \\
Summerland                & 49 36 00         &   D. Gamble            &  QHY174M-GPS        &  3                 &    Negative \\
BC, CA                    & 485              &                        & None                &                    &             \\ 
\midrule  
            & -122 15 30.0    &                        &  114                &     4              &     \\
Woodside        & 37 23 36.2      &   F. Marchis           &  eVscope/IMX347     &  4                 &    Inconclusive   \\
CA, US          & 367             &                        & None                &                    &             \\  
\midrule
Naylor Obs.               & -76 53 43.48     &  R. Kamin        &  355.6                     & 0.5                &             \\
Harrisburg                & 40 08 55.60      &  R. Young        & QHY174M-GPS                & 0.5                &  Overcast   \\
PA, US                    & 181.6            &                  &                            &                    &             \\
\midrule
Nikitin Home              & -105 11 00.96    &                  &    279.4                   &  6                &             \\
Gunbarrel                 &   40 04 08.61    &    V. Nikitin    &  QHY174-GPS                &  6                &  Overcast   \\
CO, US                    &    1,594         &                  &  None                      &                   &             \\ 
\midrule
Van Vleck Obs.            & -72 39 33.12     &  K. McGregor     &  610                       &                   &             \\
Middletown                & 41 33 18.62      &  S. Redfield     & Apogee E2V                 &  No data          &  Overcast   \\
CT, US                    & 68.9             &                  & Luminance                  &                   &             \\
\midrule
Skychariot                & -74 58 49.28     &                  & 406.4                      &                   &             \\
Shohola                   & 41 21 39.94      &  M. Sproul       & ASI1600MC                  &  No data          &  Overcast   \\
PA, US                    & 414.6            &                  & None                       &                   &             \\
\midrule
NIRo - CAC                & -87 22 31.08     &                        & 508                  &                   &             \\
Lowell                    & 41 16 15.03      &  A. W. Rengstorf       & FLI PL09000          &  No data          &  Overcast   \\
IN, US                    & 195.7            &                        & I                    &                   &             \\
\bottomrule
\end{tabular}
\end{table*}

\setcounter{table}{0}

\begin{table*}[!ht]
\centering
\caption{[continued.] Observational circumstances.}
\begin{tabular}{c c c c c c} \toprule \toprule
            & \textbf{Longitude [$^\circ$ ' '']}   &                    & \textbf{Aperture (mm)}      & \textbf{Exp. Time} &  \textbf{Status}        \\
\textbf{Site}   & \textbf{Latitude [$^\circ$ ' '']}    & \textbf{Observers} & \textbf{Detector}           & \textbf{Cycle}     & \textbf{Light curve}    \\
            &\textbf{ Altitude [m]}           &                    & \textbf{Filter}             & \textbf{(s)}       & \textbf{Dispersion}    \\ 
\midrule
Jimginny Obs.             & -88 07 0.00      &                        & 356                  &                   &              \\
Naperville                & 41 45 32.40      &   R. Dunford           & QHY174M-GPS          &  No data          &  Overcast    \\
IL, US                    & 230              &                        & None                 &                   &              \\  
\midrule
Pine Mountain Obs.        & -120 56 28.8     &   S. Fisher            & 610                  &                   &              \\
Bend                      & 43 47 30.3       &   A. Luken             & Andor Xyla           &  No data          &  Overcast    \\
OR, US                    & 1,920            &                        & None                 &                   &              \\  
\midrule  
Palomar Obs.              & -116 51 53.5     &E. Fernandez-Valenzuela & 5,100                 &                    &              \\
San Diego                 & 33 21 21.5       &   K.de Kleer           &   CHIMERA            &  No data           &  Overcast    \\
CA, US                    & 1,712            & T. Marlin, M. Camarca  &                      &                    &              \\  
\midrule  
Westport Astronomical Obs. & -73 19 39.0      &                        &  356                &                    &              \\
Westport                   & 41 10 16.0       &   K.~D.~Green          &  QHY174M-GPS        &  No data           &  Overcast    \\
CT, US                     & 87               &                        & None                &                    &              \\  
\midrule  
            & -119 09 39      &                        &  280                &                    &     \\
Yerington       & 38 59 28        &   M. Christensen       &  QHY174M-GPS        &  No data           &    Overcast   \\
NV, US          & 1,340           &                        & None                &                    &             \\ 
\midrule  
            & -121 18 55      &                        &  280                &                    &     \\
Bend            & 44 03 29        &   A-M. Eklund          &  QHY174M-GPS        &  No data           &    Overcast   \\
OR, US          & 1,105           &                        & None                &                    &             \\ 
\midrule  
            & -114 37 40      &   D. Thompson          &  280                &                    &     \\
Yuma            & 32 41 34        &    K. Conway           &  QHY174M-GPS        &  No data           &    Overcast   \\
CA, US          & 62              &   D. Conway            & None                &                    &             \\ 
\midrule  
            & -119 49 06      &                        &  280                &                    &     \\
Reno            & 39 32 42        &   T. Stoffel           &  QHY174M-GPS        &  No data           &    Overcast   \\
NV, US          & 1,387           &                        & None                &                    &             \\ 
\midrule  
            & -120 10 24      &  T. Miller             &  280                &                    &     \\
Cedarville      & 41 31 45        &  D. Schulz             &  QHY174M-GPS        &  No data           &    Overcast   \\
CA, US          & 1,420           &                        & None                &                    &             \\ 
\midrule  
            & -113 51 46      &   C. Patrick           &  305                &                    &     \\
Kingman         & 35 24 43        &   K. Butterfield       &  QHY174M-GPS        &  No data           &    Overcast   \\
AZ, US          & 966             &                        & 0.5 focal reducer   &                    &             \\ 
\midrule  
            & -116 45 29      &                        &  280                &                    &     \\
Beatty          & 36 54 04        &   J. Heller            &  QHY174M-GPS        &  No data           &    Overcast   \\
CA, US          & 999             &                        & None                &                    &             \\ 
\midrule  
            & -121 23 56      &                        &  280                &                    &     \\
Burney          & 41 02 45        &   M. von Schalscha     &  QHY174M-GPS        &  No data           &    Overcast   \\
CA, US          & 1,012           &                        & None                &                    &             \\ 
\midrule  
            & -119 44 59      &                        &  304                &                    &     \\
Gardnerville    & 38 56 29        &   J. Bardecker         & Watec 910-HX        &  No data           &    Overcast   \\
NV, US          & 1,449           &                        & None                &                    &             \\ 
\bottomrule  

\end{tabular}
\tablefoot{\footnotesize TAOS II: The Transneptunian Automated Occultation Survey. CFHT: Canada-France-Hawaii Telescope. TUHO: Tohoku University Haleakala Observatory. UCSC: University of California, Santa Cruz. NIRo: Northwest Indiana Robotic - Calumet Astronomy Center.}

\label{tab:obs-circums}
\end{table*}  
\clearpage
%%%%%%%%% Appendix B
\section{Occultation map}
\label{ap:occ_map}
\begin{figure*}
\centering
\includegraphics[width=\hsize]{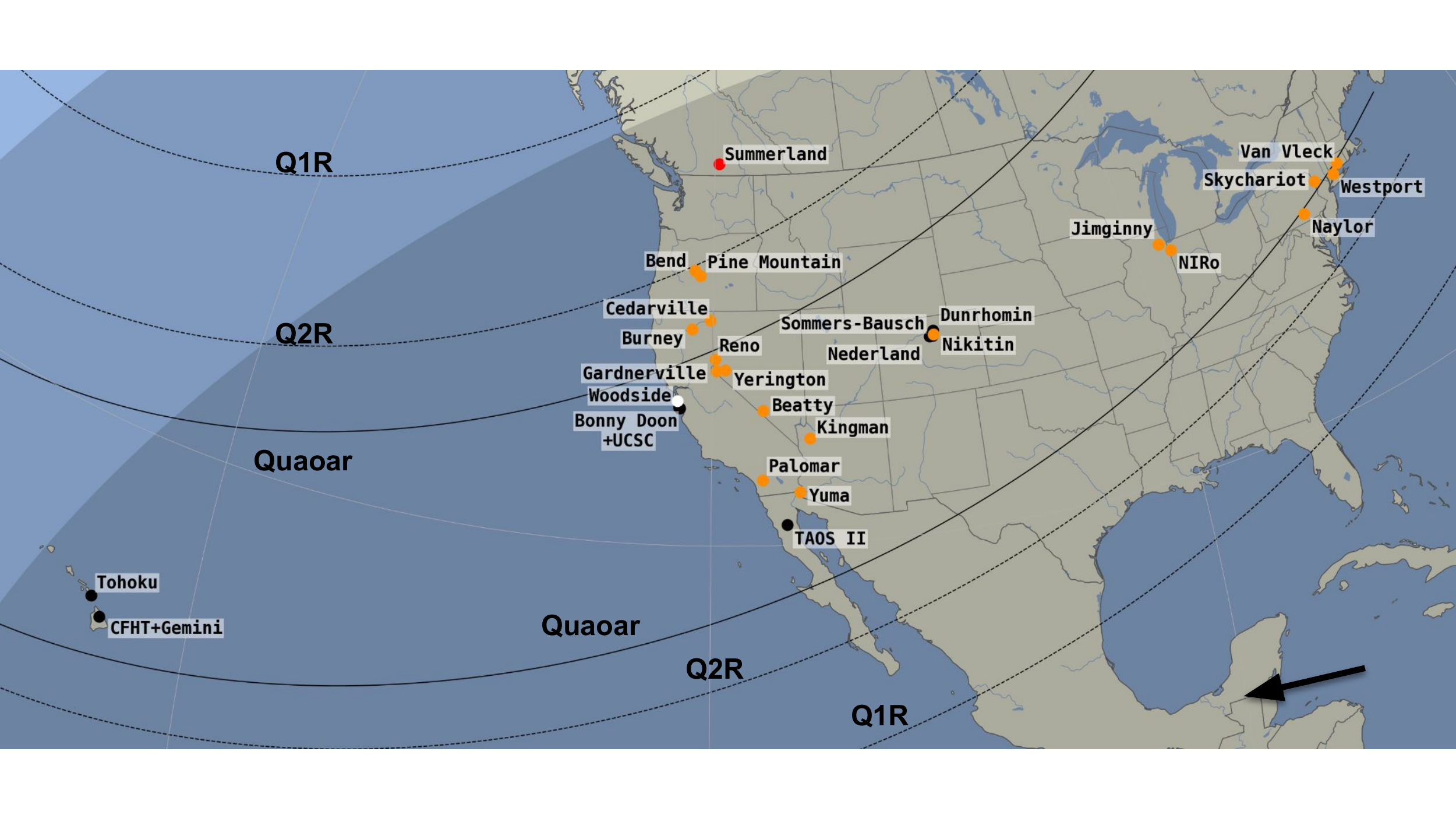}
    \caption{Occultation map with positive detections (black dots) and sites that reported cloudy weather (orange dots). The white dot represents no data due to the instrumental limits. In red we display a site with data but no occultation detected. The black solid lines mark Quaoar's shadow limits and the black dashed lines mark the Q1R and Q2R projections. The black arrow in the bottom right indicates the direction of motion of the shadow.} 
\label{fig:occ_map}
\end{figure*}

Figure \ref{fig:occ_map} presents the stellar occultation map with all the sites involved in this campaign. The green dots mark the positive detections and orange dots are stations that reported cloudy weather. The white dot represents the Woodside observation with the eVscope where the event was not detected due to the low S/N of the data set. The continuous line represents the limit of the Quaoar's shadow projected over the Earth. The dashed-lines represent the Q1R and Q2R projections, as annotated in the image.

\clearpage
%%%%%%%%% Appendix D
\section{Light curves}
\label{ap:LCs}
\begin{figure*}[!ht]
\centering
\includegraphics[width=180mm]{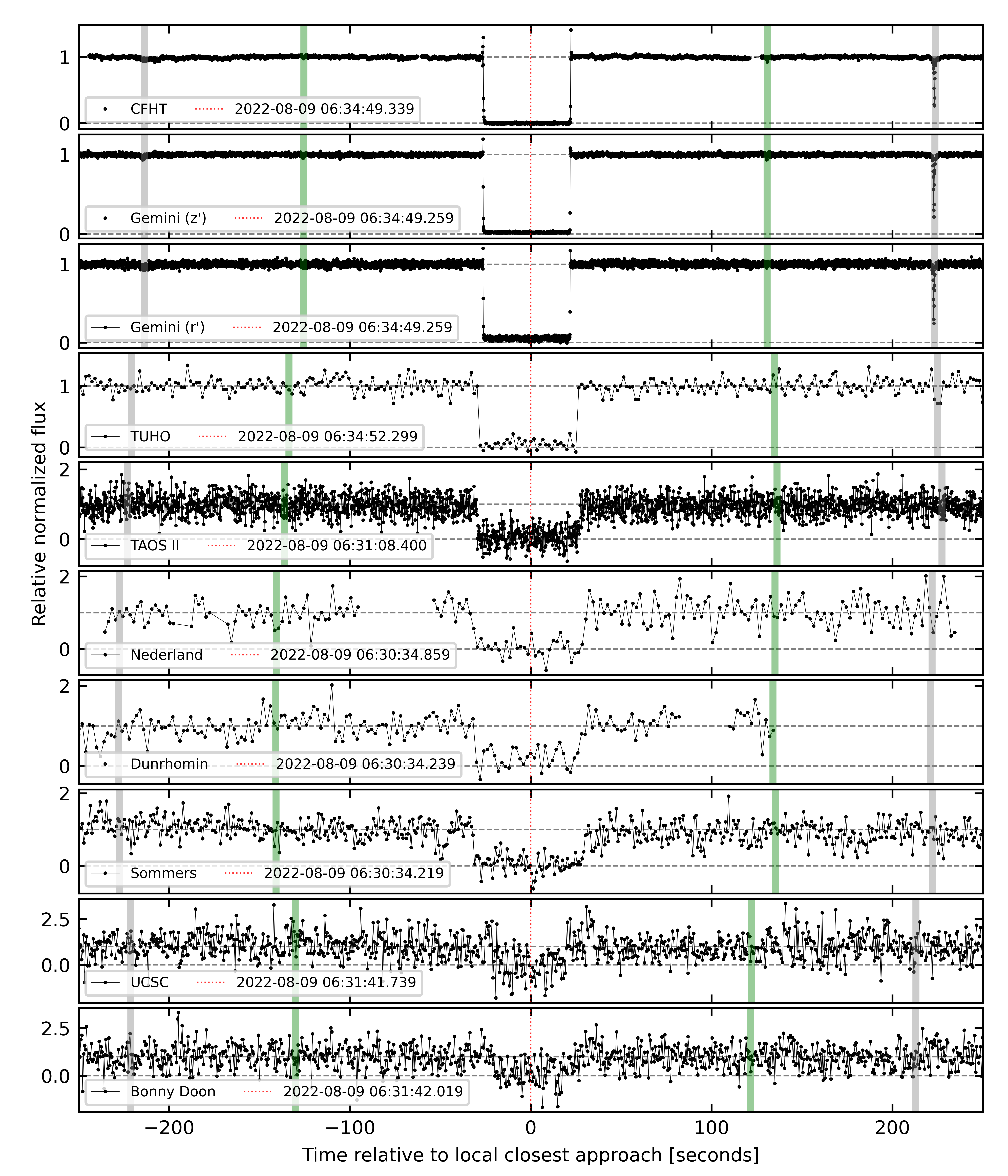}
\caption{%
All the positive light curves obtained during the August 9$^{\rm {}}$, 2022 stellar occultation by Quaoar. The black dots represent the data points. These light curves are plotted as a function of the time in seconds relative to the closest approach for each site. The green and gray vertical lines stand for Q2R and Q1R, respectively. When detected, these lines stand for calculated detection times. For light curves where these secondary structures were not detected, the green and gray lines mark theoretical times expected from our best-fit solution for a circular ring. The horizontal gray dashed lines represent the baseline and minima of the stellar fluxes.
}%
\label{fig:all_lcs}
\end{figure*}

Figure~\ref{fig:all_lcs} displays all the normalized light curves obtained during the stellar occultation by (50000) Quaoar of August 9$^{\rm {}}$, 2022. The stellar flux has been corrected for sky transparency fluctuations using reference stars. The time scale is relative to the closest approach time at each site. Besides the occultation by Quaoar and Q1R, the Gemini and CFHT light curves revealed the new Q2R ring. The Tohoku (TUHO) and TAOS II data allowed for the detection of the densest region of the Q1R ring. Due to the S/N limitations, the Nederland, Dunrhomin, Sommers-Bausch, UCSC, and Bonny Doon stations detected only the occultation by the main body. Some data sets presented dropped frames during the photometric process, such as the Durhomin and Nederland. This occurred due to the degradation of the sky quality shortly after the occultation of the main body, in addition to clouds crossing the field of view of the telescope. 

We noticed that the Gemini and CFHT light curves were misaligned in time, probably caused by an offset in the Gemini data. As we have a reliable time source for the CFHT data, we aligned the centers of these chords by applying an offset of $+\,0.27$ seconds on the Gemini chord, obtaining values of $\chi^2_{\rm{pdf}}$ closer to 1. Although very close, this result is slightly better than applying an offset to the CFHT chord.

\clearpage
%%%%%%%%% Appendix E    
\section{Modeled ring detections}
\label{ap:model_rings}
The ring events were fit using the models described in \citet{Elliot1984} and \citet{Berard2017}, that is, square boxes with uniform opacity and sharp edges. As with the main body occultation modeling, the modeled curve was convolved with observation bandpass, apparent star diameter, and instrumental response, but now fitting an opacity for the square box. The outputs of a given fit are the width, $W$, of the event, its apparent opacity, $p'$ (corresponding to the depth of the observed stellar flux drop), and its apparent optical depth, $\tau' = -\ln(1 - p')$. The geometry of the ring (assumed here to be circular) is defined by its opening angle, $B$, and the position angle, $P$, of its apparent semi-minor axis, counted positively from local celestial north to the celestial east direction. This provides the radial width, $W_{\rm r}$, the normal opacity, $p_{\rm N} = |\sin B|~(1 - \sqrt{1 - p'})$, and the normal optical depth, $\tau_{\rm N} = \tau'~|\sin B|/2$, of the ring from each event. The factor of 2 in the formula for $\tau_{\rm N}$ is due to the fact that Airy diffraction by individual ring particles results in a loss of light, resulting in an observed optical depth to be twice as large as would be measured at the ring level; see more details in \citet{Cuzzi1985, Roques1987}.
We note that the formulae for $p_{\rm N}$ and $\tau_{\rm N}$ are valid only for mono-layer and poly-layer rings, respectively. The equivalent width (respectively, depth) $E_{\rm p} = W_{\rm r} p_{\rm N}$ (respectively, $A_{\tau}= W_{\rm r} \tau_{\rm N}$) measures the amount of ring material that blocked the stellar rays in the mono-layer (resp. poly-layer) case.

A Lorentzian function is fitted to the dense part of the Q1R in the Gemini and CFHT light curves, converted from flux to normal optical depth as a function of radial distance in the ring plane. The area under the curve equals the equivalent width, A$_{\tau}$, of the ring, and the full-width half minimum (FWHM) gives us the approximate width of the ring's core. The position of the function's valley %, and its location of the peak 
gives us the average radial distance between the ring and Quaoar's center. The E$_{\rm{p}}$ value was obtained from the integral of the ring profile in the curve of the normal opacity, $p_{\rm{N}}$, as a function of the radial distance in the sky plane. This modeling considers the stellar apparent diameter of 1.33 km, which has negligible influence on the ring profile.

\begin{figure*}
\centering
\includegraphics[width=160mm]{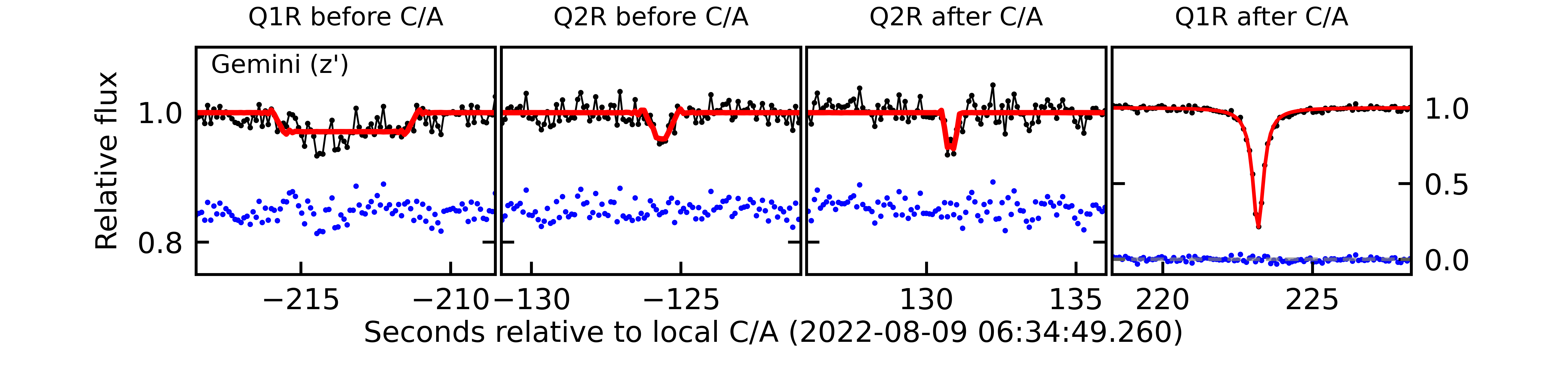}
\includegraphics[width=160mm]{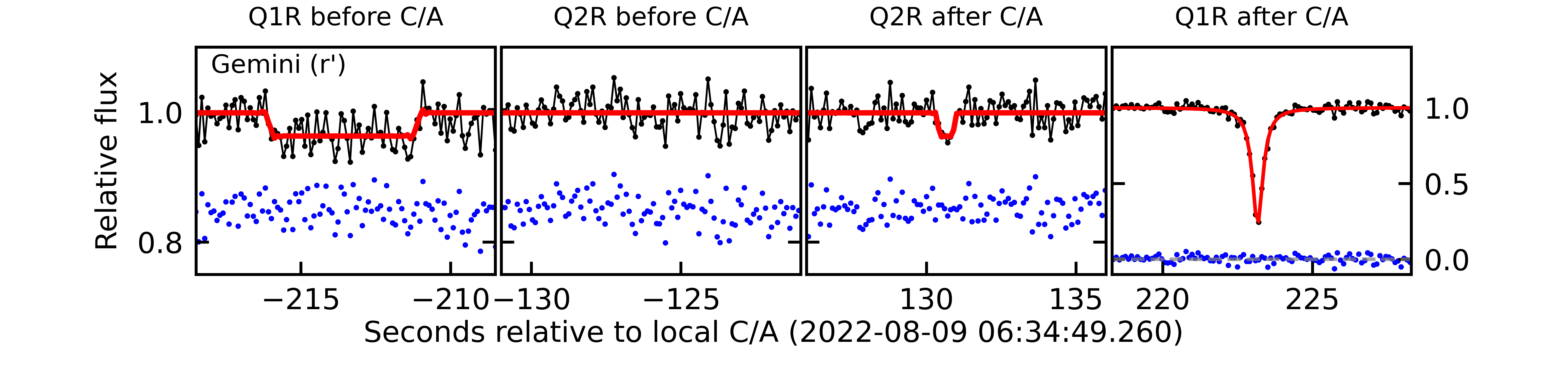}
\includegraphics[width=160mm]{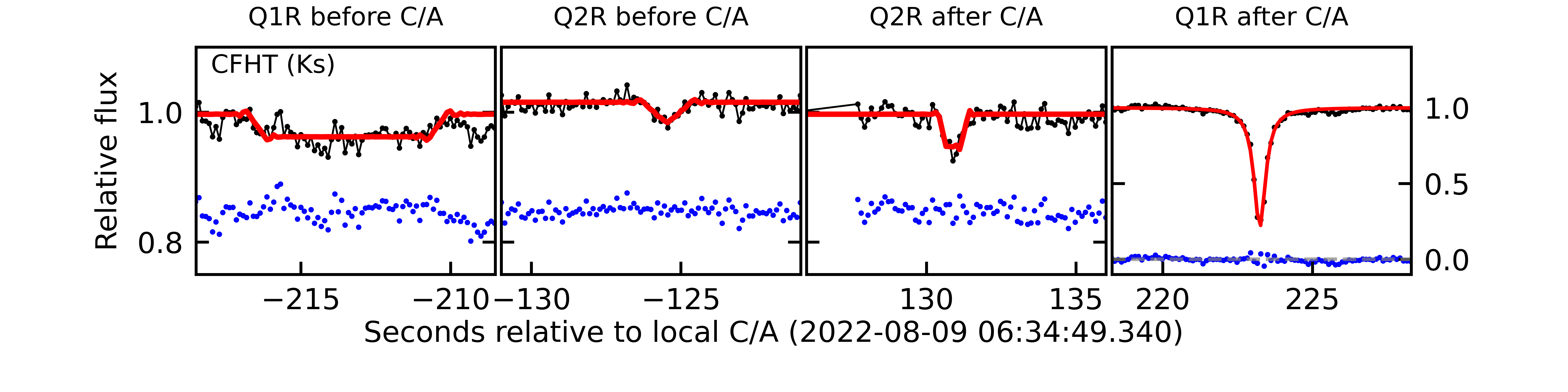}
\caption{%
Fits to the Q1R and Q2R light curve data taken with Gemini (z'), Gemini (r'), and CFHT (Ks) (the corresponding filters are indicated in parentheses). 
These light curves are plotted as a function of the time in seconds relative to the local closest approach (C/A).
The blue dots represent residuals with an arbitrary vertical offset for clarity. Note: the y-axis scale is different in the rightmost panel in each row (dense part of Q1R).
}%
\label{fig:models_1}
\end{figure*}

\begin{figure*}
\centering
\includegraphics[width=160mm]{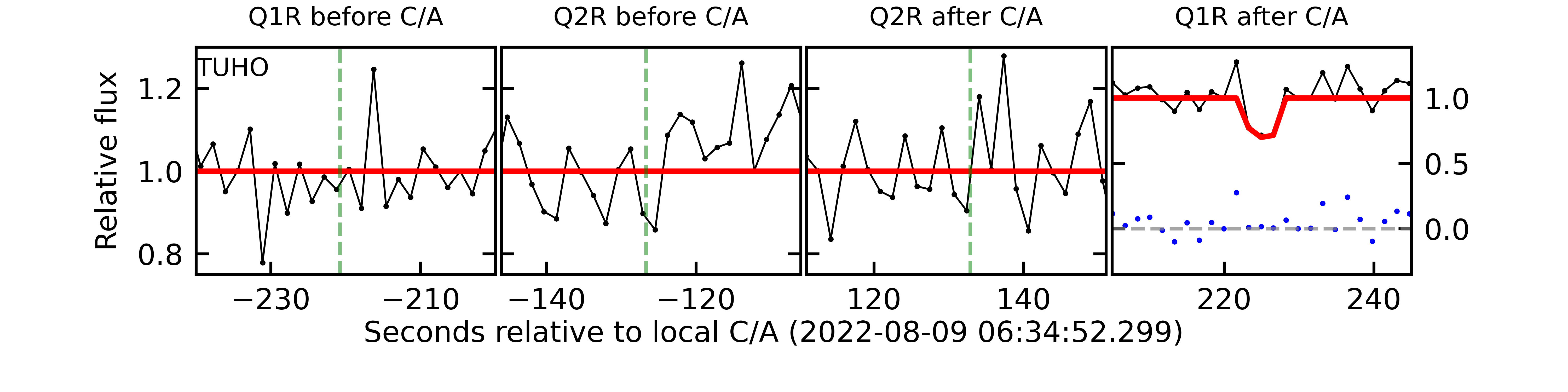}  
\includegraphics[width=160mm]{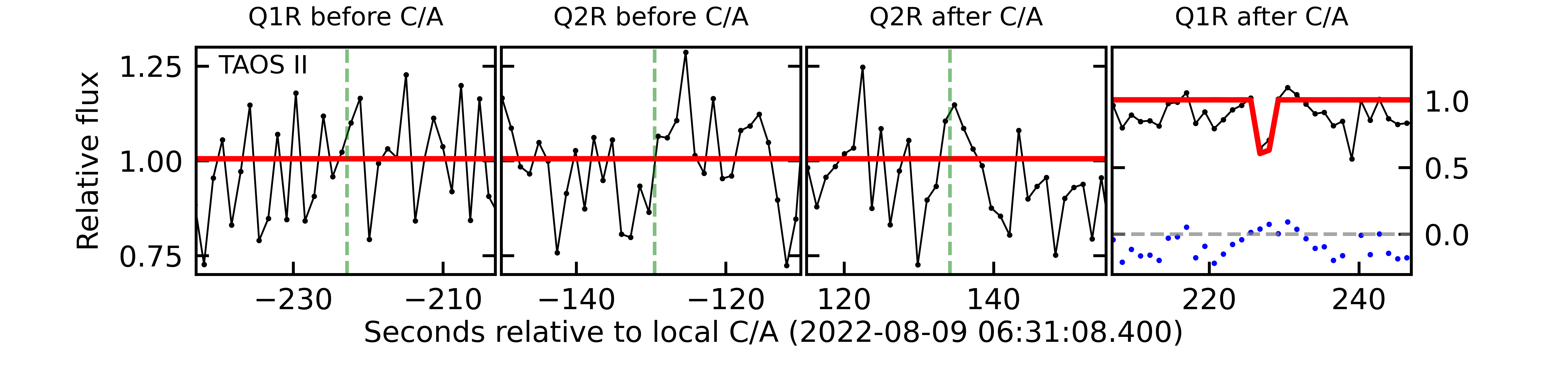}
\caption{Fits to the Q1R dense ring in light curve data taken with TUHO and TAOS II telescopes, from top to bottom. TAOS II data are stacked every six images. These light curves are plotted as a function of the time in seconds relative to the local closest approach (C/A), with a different y-axis scale in the rightmost panel in each row (dense part of Q1R).
The green vertical dashed lines represent the theoretical times for the Q1R and Q2R rings. Note: although marginal, the two data points that detect the Q1R ring in the TAOS II light curve are statistically significant at the 4.2-$\sigma$ level. 
}%
\label{fig:models_2}
\end{figure*}

Figs. \ref{fig:models_1} and \ref{fig:models_2} present all light curves in which the Q1R and Q2R rings were detected. The TAOS II light curve, used for detecting of the densest region of the Q1R ring,  was obtained by applying aperture photometry to stacked images in order to increase the S/N\bla. Stacking was performed using Python routines built from the \texttt{astropy} library as described in \cite{Morgado2019}, where each new image is the average of six original images, resulting in a temporal resolution of 1.2 seconds for each stacked image. 

\begin{table*}[!ht]
\centering
\tiny
\caption{%
Physical parameters of rings Q1R and Q2R.
}%
\begin{tabular}{l l c c c c c c c} \toprule \toprule
\multirow{2}{*}{Ring}   & \multirow{2}{*}{Detection} & Mid-time & $r$    &   $W_{\rm r}$    & \multirow{2}{*}{$p_{\rm N}$}  &  \multirow{2}{*}{$\tau_{\rm N}$}  &   $\rm{E}_{\rm p}$  & $\rm{A}_{\tau}$ \\
                        &                       & August 9$^{\rm {th}}$, 2022      &(km) &   (km)     &                       &                           &   (km)   & (km)       \\
\midrule
\multirow{3}{*}{Q1R- ing}    & CFHT           & 06:31:15.8 (0.2)        & 4,009.9 (1.8)           & 105.5 (5.6)   & 0.0030 (0.0003)       & 0.0030 (0.0003)       & 0.31 (0.02)        & 0.31 (0.02) \\ 
                             & Gemini (z')    & 06:31:15.72 (0.04)      & 3,995.4 (0.3)         & 76.4 (0.8)      & 0.0026 (0.0001)       & 0.0026 (0.0001)            & 0.20 (0.01)           & 0.20 (0.01) \\ 
                             & Gemini (r')    & 06:31:15.7 (0.1)        & 3,995.9 (0.8)         & 85.8 (1.1)      & 0.0032 (0.003)            & 0.0032 (0.0003)       & 0.28 (0.02)        & 0.28 (0.02) \\\cmidrule(r){1-9}
\multirow{5}{*}{Q1R- egr}   & CFHT$^a$          & 06:38:32.581 (0.003)       & 4,123.69 (0.05)       & 6.1 (0.1)     & 0.330 (0.007)    & 0.395 (0.009)         & 2.01 (0.03)       & 2.41 (0.04) \\
                            & Gemini (z')$^a$   & 06:38:32.444 (0.003)       & 4,123.11 (0.02)       & 5.3 (0.1)     & 0.353 (0.008)    & 0.417 (0.007)         &   1.87 (0.02)     &  2.21 (0.02) \\
                            & Gemini (r')$^a$   & 06:38:32.429 (0.005)       & 4,122.85 (0.04)       & 5.1 (0.1)     & 0.341 (0.009)    &  0.398 (0.008)        &   1.74 (0.03)     &  2.03 (0.03) \\
                            & TUHO              & 06:38:37.4 (0.7)           & 4,107.2 (7.6)          & 71 (17)       & 0.03 (0.01)          & 0.03 (0.01)          & 2.3 (0.9)             & 2.4 (1.1)) \\ 
                            & TAOS II           & 06:34:55.8 (0.5)           & 4,125.7 (4.4)          & 34 (16)       & 0.05 (0.02)          & 0.05 (0.03)          & 1.6 (1.1)             & 1.7 (1.3) \\
\midrule
\multirow{3}{*}{Q2R- ing}   & CFHT              & 06:32:43.9 (0.2)           & 2,490.4 (1.6)          & 16.5 (3.0)        & 0.0022 (0.0004)      & 0.0022 (0.0004)     & 0.036 (0.006)         & 0.036 (0.006) \\
                            & Gemini (z')       & 06:32:43.61 (0.06)         & 2,493.8 (0.6)         & 11.3 (0.8)         & 0.0034 (0.0005)  & 0.0034 (0.0005)            & 0.038 (0.006)         & 0.039 (0.006) \\
                            & Gemini (r')           & n.d.$^b$                   & n.d.$^b$              & n.d.$^b$      & n.d.$^b$         & n.d.$^b$              & n.d.$^b$          &  n.d.$^b$ 
                            \\\cmidrule(r){1-9}
 \multirow{3}{*}{Q2R- egr}  & CFHT              & 06:37:00.20 (0.01)         & 2,540.7 (0.1)          & 10.8 (0.2)        & 0.0049 (0.0003)  & 0.0049 (0.0003)            & 0.052 (0.001)         & 0.053 (0.001)  \\    
                            & Gemini (z')       & 06:37:00.09 (0.02)         & 2,540.45 (0.1)        & 6.3 (0.3)  & 0.0047 (0.0005)  & 0.0048 (0.0006)       & 0.030 (0.002)          & 0.030 (0.002) \\
                            & Gemini (r')       & 06:36:59.93 (0.02)         & 2,528.3 (1.1)         & 9.1 (2.5)  & 0.003 (0.001)        & 0.003 (0.001)         & 0.030 (0.008)      & 0.030 (0.008)  \\
                            \bottomrule
\end{tabular}
\tablefoot{The normal opacity, $p_{\rm N}$, and normal optical depth, $\tau_{\rm N}$, were calculated from the ring opening angle, $B$, and position angle, $P$, on August 9$^{\rm {}}$, 2022, derived from the orientation of the body obtained from the 2018-2022 data. The other parameters are: Mid-time in UT and the error bars in parentheses given in seconds\bla; $r$, the radial distance from Quaoar's center in kilometers; $\rm{W}_{\rm r}$, the radial width in kilometers; $\rm{E}_{\rm p}$ and $\rm{A}_{\tau}$, the equivalent width and equivalent depth in kilometers respectively. The terms ``ing" and ``egr" stand for ingress and egress, respectively, and refer to the fact that the detection occurred before and after the occultation by Quaoar's main body, respectively. The error bars in parentheses are at the 1-$\sigma$ level. $^a$From Lorentzian fit. The width $W_{\rm r}$ is defined as the FWHM of the $\tau_{\rm{N}}$ profile. The E$_{\rm{p}}$ and A$_{\tau}$ values where obtained from respective integrals in the ring profile (see Sec. \ref{sec:data_analysis}); $^b$Not detected.}
\label{tab:rings_local}
\end{table*}

\end{appendix}
\end{document}